\RequirePackage{fix-cm}
\documentclass[referee]{svjour3}                     
\smartqed  
\usepackage{graphicx}
\usepackage{a4wide}
\usepackage{natbib}
\journalname{Photosynthesis Research}
\begin{document}

\title{Constrained geometric dynamics of the Fenna-Matthews-Olson complex
}
\subtitle{The role of correlated motion in reducing uncertainty in excitation energy transfer}


\author{Alexander S. Fokas         \and
        Daniel J. Cole \and
        Alex W. Chin
}


\institute{A.S. Fokas \at
              The Theory of Condensed Matter Group Group, Cavendish Laboratory, 19 JJ Thomson Avenue, CB3 0HE, Cambridge, U.K. \\
              Tel.: +44(0)1223-337460\\
              \email{asf40@cam.ac.uk}           
           \and
           D.J. Cole \at
              The Theory of Condensed Matter Group Group, Cavendish Laboratory, 19 JJ Thomson Avenue, CB3 0HE, Cambridge, U.K.\\
              Department of Chemistry, Yale University, 225 Prospect Street, New Haven, Connecticut 06520-8107, United States\\
              \email{daniel.cole@yale.edu}
          \and
          A.W. Chin  \at 
             The Theory of Condensed Matter Group Group, Cavendish Laboratory, 19 JJ Thomson Avenue, CB3 0HE, Cambridge, U.K.\\
             \email{ac307@cam.ac.uk}
}

\date{Received: date / Accepted: date}

\maketitle

\begin{abstract}
The trimeric Fenna-Mathews-Olson (FMO) complex of green sulphur bacteria is a well-studied example of a photosynthetic pigment-protein complex,  in which the electronic properties of the pigments are modified by the protein environment to promote efficient excitonic energy transfer (EET) from antenna complexes to the reaction centres. By a range of simulation methods, many of the electronic properties of the FMO complex can be extracted from knowledge of the static crystal structure. However, the recent observation and analysis of long-lasting quantum dynamics in the FMO complex point to protein dynamics as a key factor in protecting and generating quantum coherence under laboratory conditions. While fast inter-- and intra-- molecular vibrations have been investigated extensively, the slow, conformational dynamics which effectively determine the optical inhomogeneous broadening of experimental ensembles has received less attention.  The following study employs constrained geometric dynamics to study the flexibility in the protein network by efficiently generating the accessible conformational states from the published crystal structure. Statistical and principle component analysis reveal highly correlated low frequency motions between functionally relevant elements, including strong correlations between pigments that are excitonically coupled. Our analysis reveals a hierarchy of structural interactions which enforce these correlated motions, from the level of monomer-monomer interfaces right down to the $\alpha$-helices, $\beta$-sheets and pigments. In addition to inducing strong spatial correlations across the conformational ensemble, we find that the overall rigidity of the FMO complex is exceptionally high. We suggest that these observations support the idea of highly correlated inhomogeneous disorder of the electronic excited states, which is further supported by the remarkably low variance (typically \textless  5 \%) of the excitonic couplings of the conformational ensemble.
\end{abstract}
\section{Introduction}
\label{intro}

The early stages of photosynthesis involve ultrafast photophysics, which facilitate the capture, transport, and, ultimately, utilisation of photonic energy. Thereafter, photosynthetic organisms are able to initiate a much slower sequence of complex chemical events that are necessary for life. These initial stages, known as the light reactions, employ optically active molecules (pigments) tuned to the organism's photonic environment. Light-harvesting proteins have been developed in Nature for the purpose of co-ordinating the pigments and other cofactors. They are employed for a variety of different functions, such as energy transporting antennas, charge separation centres, photoprotection mechanisms, and soforth \citep{blankenship,van2000photosynthetic}. The versatility of the pigment-protein organisational paradigm has enabled photosynthetic organisms to colonise a diverse array of environments, and there is currently a great interest in learning how the reproducible control and efficiency of nanoscale energy and charge processes in natural pigment-protein complexes (PPCs) is realised \citep{scholes2011lessons}. 

Part of the motivation for this is the rapid advance of nanoscale materials engineering in the field of solar light harvesting, particularly in organic optoelectronics, where biological solutions to problems of efficient and directed excitation energy transfer (EET) and stable, long-range charge separation could suggest new ways to break these otherwise limiting constraints in artificial devices \citep{scholes2011lessons,dimitrov2013materials}. Moreover, the scope of what may be learned from PPCs has recently been expanded due to the unexpected observation of room temperature quantum mechanical dynamics in a range of PPCs extracted from bacteria, plants and algae \citep{lee2007coherence,harel2012quantum,collini2010coherently,277K,engel2007,robust_energy_disorder}. The potential advantages of coherent quantum
dynamics in energy transfer include, {\it inter alia}; robustness against disorder, enhanced rates via interference (constructive and negative), quantum thermal ratching, and lower driving forces for directed transport~\citep{good_review,robust_energy_disorder,mohseni2008environment,plenio2008dephasing,caruso2009highly,huelga2013vibrations,chin2013role,fassioli2014photosynthetic,QB}.  Although direct, unambiguous evidence for the beneficial role of quantum effects in photosynthesis remains elusive, an experimental example of how transient coherence can be crucial in organic photophysics has recently been presented by \citet{gelinas2014ultrafast}, where ultrafast ($<80$~fs) electron-hole separation at a bulk heterojunction only occurred when well-ordered acceptor phases which support delocalised electron states were present. This ultrafast, coherent electron dynamics leads to sufficiently rapid charge separation which can overcome the strong Coulomb binding energy that otherwise traps charges close to the heterojunction interface and drastically reduces their internal quantum efficiency \citep{gelinas2014ultrafast}. As has been understood for a long time,  the structure and motion of the protein environment is fundamentally important for understanding how quantum dynamics might impact on EET efficiency, and a nascent interdisciplinary field, known as quantum biology, has subsequently begun to explore the surprising stability and role of quantum coherence in these structures. Acquiring a better understanding of these properties                                                                          could provide interesting insights for other emerging types of quantum technology \citep{QB,huelga2013vibrations,anna2014little,scholes2011lessons}.

Though PPCs are ubiquitous
in photosynthetic organisms, much of the attention related to quantum dynamics has focussed on the Fenna-Matthews-Olson
(FMO) complex of green sulphur bacteria (GSB) \citep{van2000photosynthetic,blankenship}. This structure facilitates the transport of electronic (excitation) energy from the antenna super-complexes to the reaction centres (RCs). \citet{engel2007} employed ultrafast femtosecond non-linear optical experiments to observe long-lasting (ps) quantum beating between electronic excited states in a four-wave-mixing response of an ensemble of FMO proteins. Although it is only found in GSB, the FMO complex carries out a typical and important function that is found in all the light reactions: namely, EET \citep{van2000photosynthetic,blankenship}. The EET dynamics are driven by the dipole-dipole coupling of the pigments' optically excited states and the presence of a spatial energy gradient of local pigment excitation energies \citep{van2000photosynthetic,blankenship,Adolphs:2006yq,direct}. The latter promotes irreversible energy relaxation that causes excitations to move to a specific spatially localised site or complex. It has been demonstrated that the FMO protein has evolved to drive the excitation close to the RC where this energy initiates the primary photochemical event of photosynthesis; charge separation \citep{mem_orient}. 

As is now well-established, efficient EET dynamics depend on several key electronic parameters which are \textit{controlled} by the pigment-protein structure; namely, (a) the local pigment optical transition
energies (site energies), (b) the couplings between particular pigment
optical transitions (excitonic couplings), and (c) the dynamical spectral density of protein fluctuations,
which describes the strength and timescale of site energy modulations caused by the
dynamic, thermal motion of the pigments' molecular surroundings and internal structure. This latter property provides the dissipative element required to drive relaxation along the excitation energy gradient. Many previous theories of EET in FMO have employed models which incorporate these parameters to varying extents, allowing the prediction of kinetic and optical responses during EET. However, the necessity to include quantum coherence in the EET description introduces the need for new and sophisticated real-time simulations of density matrix evolution, higher precision prediction of key molecular parameters from {\it ab initio} techniques, and a reappraisal of the electronic and protein structure in terms of its capacity to host and coordinate quantum effects \citep{QB,huelga2013vibrations,anna2014little,scholes2011lessons}.

In this paper, we aim to provide an atomistic survey of the properties of the FMO protein, exploring the conformal motion and ensemble of its realisable structures using recently developed tools for analysing very large molecular systems. In our focus at all times will be the consequences of the multiscale, hierarchal structural relationships our simulations reveal, which we suggest facilitate coherent EET and aid interpretation of the  experimental data. In this latter respect, a key result we shall present is the reduced disorder of the internal pigment arrangements from realisation to realisation. We suggest this property may act to suppress the inhomogeneous dephasing between excited states. This quality of the system is made possible by the highly coordinated  motion of a hierarchy of structural elements in the FMO. This might offer some support for correlated-disorder theories that have attempted to explain the much weaker than expected effect of the substantial inhomogeneity observed in the ground state absorption of FMO on the longevity of inter--excited state coherences \citep{fidler2011two}. Moreover, we also speculate that the correlations of structural elements within \emph{each} realisation might act to stabilise the static, average Hamiltonian structure of each configuration against inevitable deformations, helping the ensemble to perform efficient EET by ensuring that most of the members have functional electronic structure (see Section \ref{sec:dissipative}).

The structure of this paper is as follows: The key role of the protein structure in tuning (a)--(c), and how this affects quantum mechanical EET will be developed in Section~2. We will then present our methods for structural simulation and gathered results, which allowed us to visualise the conformational motion, spatial correlations, and a hierarchy of structural and dynamical relationships found in the FMO complex. Importantly, all of these observations appear to dramatically stabilise the internal co-ordination of the `working parts' of this molecular wire.  These are then analysed in molecular detail in Section~5, after which we discuss a number of general organisational principles that act to stabilise EET in the presence of the slow protein motion that gives rise to conformal disorder. The possible relevance of the structure and correlated conformational motion for quantum dynamics and ultrafast optical experiments is highlighted, and we conclude with a few speculations on other types of quantum effect that might enable the \emph{complete trimer structure} of FMO to avoid failures in EET transmission.

\section{Pigment-protein crystal structure and an overview of its role in EET}

\begin{figure*}
\includegraphics[width=1.0\textwidth]{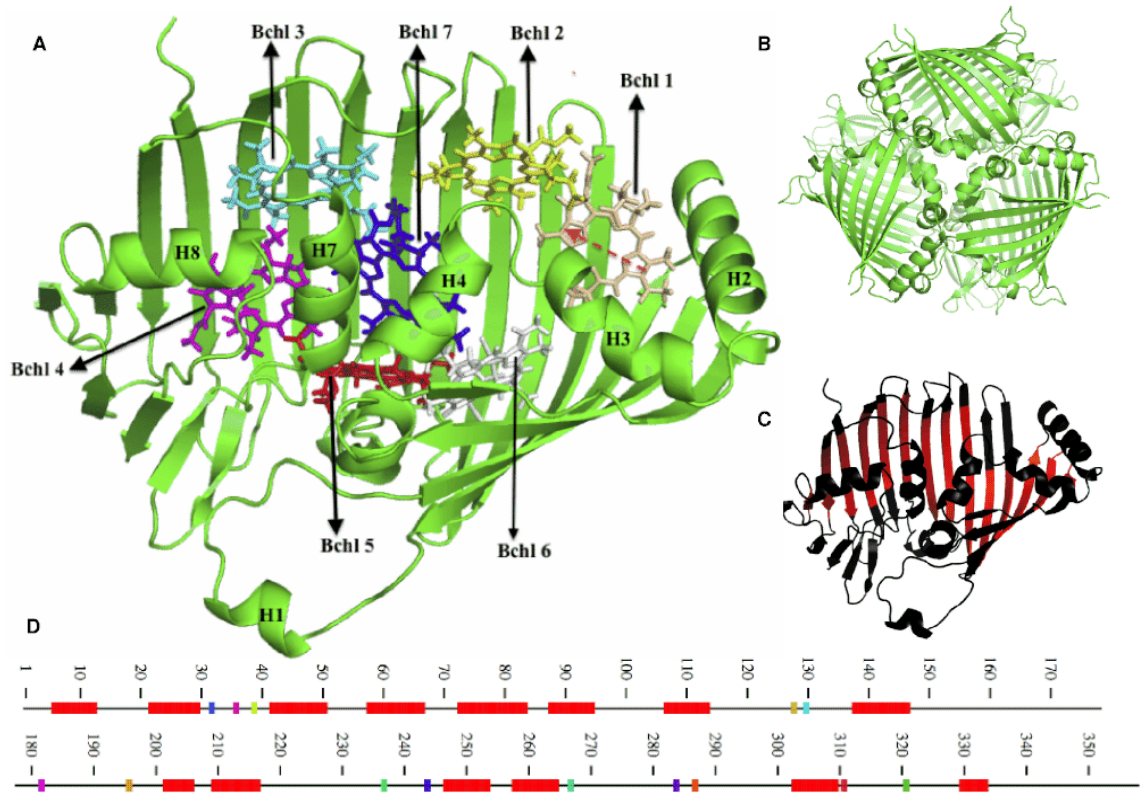}
\caption{ (A) Monomer 1 of the FMO complex with the Bchl $1$ to $7$
  and  $\alpha$-helices 1-5, 7, and 8 labeled. Bchl $1$ has
  been used to illustrate the Q$_{y}$ axis, which lies along the line formed
  by connecting the nitrogen atoms of pyrrole I and pyrrole III (red arrow). (B) The
  orientation of each monomer in the full trimer structure. The RCs 
  identified by FIRST  using $H_{cut} = -4.6$~kcal/mol are shown in red and flexible
  regions in black. These regions have been visualised on the monomer structure (C) 
  and along the primary sequence (D).}
\label{fig1}
\end{figure*}

The structure of the FMO complex is shown in Fig. \ref{fig1}. The FMO complex possesses a trimer quaternary structure in which each monomer assumes
a clam-like architecture \citep{cath} that sequesters seven pigments
\citep{tronrud}. The orientation of the FMO complex in the membrane
positions an eighth pigment, which is located on the outside of each
FMO monomer, near the baseplate (which is connected to the much larger light harvesting antenna complex known as the chlorosome) \citep{mem_orient}. The main low-energy optical absorption transition
undergone by Bacteriochlorophyll a (Bchl) is the Q$_{y}$ transition,
which has an associated transition dipole moment that lies along the
y-axis of the pigment (see Bchl 1 in Figure \ref{fig1}). The localised excited state created by either optical absorption or resonant energy transfer from another complex forms the basis of energy movement through the complex. This transport is thought to connect excitations accepted from the antenna complexes at Bchls $6$ or $1$ (from the baseplate or, possibly, Bchl $8$) and move them to Bchl $3$ \citep{mem_orient,Adolphs:2006yq,direct}.

\subsection{Site energies}
The effect of the protein on the site energies of the pigments has been previously investigated analytically using the static
crystal structure \citep{Adolphs:2006yq,direct,Adolphs:2008fu,QMMM_env}. The protein environment was found to modify the
site energy of the pigments in a number of ways. Hydrogen bonds formed between the protein and the pigments are found always
to red shift the site energy, while the electrostatic dipole moment of
the $\alpha$-helices are found to affect the transition density in a manner that depends on the relative orientation of the pigments in the protein matrix. For example, Bchl $3$ displays a site energy shift of $-103$ cm$^{-1}$ due to a hydrogen bond with Tyr $15$ and $-145$ cm$^{-1}$ due to the electrostatic potential of $\alpha$-helix 7 \citep{Adolphs:2008fu}. Similar effects of the protein on the local pigment $Q_{y}$ site energies have also recently been found in fully-quantum mechanical simulations by  \citet{cole2013}. Interestingly, QM/MM calculations have shown that Bchl $3$ exhibits the highest energy in the absence of the environment \citep{QMMM_env}. This finding completely contradicts experimental observations, which have shown it to behave as an energy sink. Therefore, the presence of the protein determines the energy landscape of the pigments to create an energy
funnel that relaxes excitation energy irreversibly toward the reaction
centre \citep{mem_orient,Adolphs:2006yq,direct}. Maintaining this energy
funnel at ambient temperatures, thereby avoiding the
the formation of new, inefficient local minima in `poor' conformations, 
appears to be enforced by a seemingly unique hierarchal structure which we shall describe.  

\subsection{Excitonic couplings}
The spatial co-ordination of the pigments on the 10 --100 nm scale of typical PPCs is of crucial importance, as the radiative lifetime of a similar concentration of Chlorophyll-based pigments in solution would be far too short (due to concentration quenching) to transfer energy effectively \citep{quench}.    
The excitonic, or resonance energy transfer, coupling that occurs between
pigments results from the Coulombic interaction between their transition dipole moments \citep{van2000photosynthetic,Adolphs:2006yq}. In the absence of environmental interactions, this leads to the coherent (wave-like) spreading of an excitation over the pigments. Quantum mechanically, this can be understood as the formation of new excited states of the system which are coherently delocalised over the pigments \citep{van2000photosynthetic,Adolphs:2006yq}. The strengths of these couplings are dependent on the mutual orientation and separation of the molecular transition dipoles. The  degree of delocalisation of the excitonic states is a sensitive function of the strength of dipole-dipole couplings and the differences in site energies of the pigments involved, such that coupling strengths significantly greater than the energy differences leads to fully delocalised excitons. In the FMO complex, the crystal structure ensures that the most significant couplings are those between nearest-neighbour pigments, creating coherently delocalised excitations. Due to the typically larger site energy differences between neighbouring pigments (compared to the excitonic couplings), this leads to excitations coherently delocalised over 1 -- 3 pigments. These new `good' quantum states, usually referred to as excitons, have modified optical properties and transition energies, the tuning of which could also be used to lend directionality to energetic relaxation dynamics. In 2D Fourier transform spectroscopies, broadband laser pulses are able to excite superpositions of these excitons. The coherent optical interference is manifested as the quantum beats (discussed in the introduction) and the ps decay times are thought to indicate sustained quantum coherence in the EET process. However, if these effects are important for function, the FMO structure must be robust against any major variations in the relative pattern of the pigments' positions and orientations. Indeed, we shall show that the excitonic couplings, treated in the point-dipole approximation, are extremely robust ($<5\%$) with respect to the allowed conformal variations supported by the FMO structure.      

\subsection{Dissipative protein fluctuations}
\label{sec:dissipative}

The formation of an energetic ladder of delocalised excitations does not in itself lead to efficient EET, as a source of dissipation and dephasing is also required to force relaxation of the excitons \citep{van2000photosynthetic,Adolphs:2006yq,plenio2008dephasing,caruso2009highly,mohseni2008environment}. This is provided by the excitable vibrational fluctuations of the pigments, protein and solvent environments which couple to the presence of locally excited electronic states, which is typically described using linear displaced oscillator models. These dynamical local interactions compete with the formation of delocalised states and cause dephasing between any prepared superposition of them. The strength and typical timescales of these fluctuations determines the qualitative character of the EET process. The kinetics fall within two extremes, namely; spatially localised, incoherent rate kinetics (F\"{o}rster theory) and kinetics in the space of delocalised states (secular Redfield theory). This interpolation of behaviour is associated with strong and weak coupling relative to the strength of excitonic couplings, respectively. In both extreme cases, the existence of \textit{inter-exciton} coherence does not impact on the EET kinetics. It is increasingly appreciated that many PPCs have roughly balanced excitonic and dynamical environmental interactions. This regime, where coherences may play a significant, albeit transient, role in EET, is less well understood. Exploration of open quantum dynamics in the  ``intermediate'' regime, particularly with highly structured environmental spectra, is currently an area of intense activity \citep{christensson2012origin,kreisbeck2012long,tiwari2013electronic,chin2013role,o2014non,intermediate_rev}.

The environmental interactions that influence EET dynamically are fast, intra-- 
and inter--molecular vibrations (10--1000s fs periods), typically of frequencies close to 
the energy differences between excitons. This is necessary for energy relaxation, as the phonons must be capable of absorbing these energies for the excitons to effectively relax. Classical molecular dynamics (MD) has been used to simulate atomistic motion on a ns timescale, but cannot be used to investigate
important optical properties of PPCs, such as the effect of motions on the excited state site energies. MD simulations have therefore
been used in combination with quantum chemistry calculations to investigate
the excited states of the system. For example, ZINDO/S calculations have been applied to MD-generated snapshots to study (a) the electronic-vibrational couplings in the Q$_{y}$ states of the pigments and (b) the transition
energy fluctuations that are required to compute the environmental spectral
density. This coupled QM/MM method has been applied to the RC of \textit{Rb. sphaeroides} \citep{QMMM_rb}
and the FMO complex of \textit{C. tepidum} \citep{spectral_dens}.

Unfortunately, the MD trajectory was unable to demonstrate the effect
of long timescale structural changes (static disorder) on the exciton
dynamics. This resulted in spectral features that are narrower than
observed experimentally, as very low frequency contributions to the fluctuation spectrum provide the main source of pure dephasing processes and inhomogeneous broadening \citep{kreisbeck2012long}.  Additionally, using a related methodology
to that described above, it was concluded that there is no correlation
in environment-induced site energy fluctuations \citep{correlation} on femto-to-nanosecond timescales.
This seemingly rules out the proposal that quantum
coherences could be protected by dynamical correlation of site-energy fluctuations \citep{fluctuation_correlation}. However,
the MD methodology  used to simulate structural changes in the environment
could only extend over 10 ns. It follows that some important characteristics
could lie hidden in the ms timescales of static disorder that occur
in natural biological systems. 

With the inability of MD to account for relatively slow conformational dynamics of the entire FMO complex, the calculation of transition
energies and their couplings to intra-- and inter--molecular motion
of the sluggish protein environment is a challenging task. Nonetheless, some progress has
been made in this area. 
Normal mode analysis (NMA) has been employed to overcome the sampling
efficiency limitation that is met when employing MD simulations. In
particular, NMA has been used to study the intermolecular exciton-vibrational
couplings, which subsequently allows information about the spectral
density to be obtained \citep{NMA_protein}. However, although the low-frequency
portion of the spectral density can be computed directly, there is
no information about which of the normal modes are the most functionally,
and therefore biologically, relevant \citep{NMA_prob}.

In this study, the slow, conformational dynamics of the FMO complex
will be simulated using the FRODA software, which takes as input
geometric constraints that are calculated by the FIRST software
\citep{bath32820} based on the experimentally derived  structural properties of
the protein.
FRODA has been widely used, for example, in studying the impact of
intracellular flexibility upon the conductance and gating of the 5-HT3
and nictotinic acetylcholine receptor ion
channels~\citep{kozuska2014,belfield2014}, protein-protein docking
involving multiple conformational changes~\citep{jolley2006},
elucidating cisplatin cross-linking in calmodulin~\citep{li2012}, and
monitoring flexibility of myosin during the ATPase
cycle~\citep{sun2008}.
In contrast to force field based molecular mechanics simulations, the constraints-based Lagrangian dynamics that is implemented in FRODA effectively flattens the potential energy surface and substantially reduces the number of degrees of freedom to be explored. It has been shown that these simplifications actually allow FRODA to outperform MD in the sampling of transient pockets at protein-protein interfaces \citep{metz2011}. Thus, the methods used here potentially allow the computationally efficient exploration of conformational space of the FMO complex, thus providing a tool to study the dynamical nature of the protein and to test theories concerning relationships between correlations and function, which complements information gained from the static crystal structure.
Importantly, the resulting structural ensembles generally represent
low frequency motions that are not necessarily small and thus go
beyond the linear (harmonic) approximation, which is assumed when
accounting for environmental dynamics by a spectral density, or normal
mode analysis. 

Therefore, by studying the concerted and correlated motions in the PPC, we hope to generate an all-atom picture of the biologically relevant motion present in the full FMO complex and thereby shed new light on the mechanisms that support EET and how long-lived quantum beats may appear in ensemble experiments. However, we note here that the motion that we analyse potentially takes place on much longer timescales than EET, and is best thought of as a computational approximation to the configurational ensemble of complexes that is found in any experiment (which is the principle contribution to inhomogeneous broadening). Spatial correlations of site energies due to this motion, if present, could thus dramatically reduce the inhomogeneous broadening and might also help maintain the relative energy level structure, known to generate efficient EET, of the crystal structure in individual members of the ensemble (preventing, for example, the appearance of configurations where site 3 is not a sink, etc.). However, such effectively static correlations on the timescale of EET are not expected to lead to any dynamical protection of coherences.

\section{Methods}
\label{sec:1}

Calculations were performed on the holo form of the trimeric 1.3~\AA{}
X-ray crystal structure of \textit{Prosthecochloris Aestuarii} (PDB
accession code 3EOJ).
Full details of the structure preparation are described
elsewhere~\citep{cole2013}.
Briefly, hydrogen atoms were added to the structure using the
MOLPROBITY software.
A manual investigation of the structure resulted in the transfer of a
proton from the $\epsilon$ to the $\delta$ nitrogen of His $6$ (and
the equivalent residues in monomers $2$ and $3$). The hydrogen atom of
Tyr $9$ and Tyr $338$ sidechains was also rotated to form hydrogen bonds
with Bchl $3$ and $4$, respectively.
The AMBER11 software~\citep{AMBER11} was employed for structural
relaxation.
The description of the protein was accomplished using the FF99SB force
field, the water molecules using the TIP3P model, and the pigments by
the force field developed by~\citet{force_field_pig}.
Missing heavy atoms were added using the {\it leap} module of AMBER11.
The structure was solvated and heated to 300~K over a period of
300~ps, followed by 1000~steps of conjugate gradients minimisation.
Strong restraints (1000~kcal/mol/\AA{}$^2$) were applied to the heavy
atoms of the protein and pigments throughout the equilibration
procedure.
A number of water molecules were found to play important roles in the
pigment-protein hydrogen bonding network.
Therefore, at the end of the equilibration procedure, the closest 600
water molecules to the pigments were retained to model the effects of
specific pigment-water and protein-water hydrogen bonds.

Constrained geometric simulations were performed using the FIRST/FRODA
software (version~6.2.1, downloaded from
http://flexweb.asu.edu) ~\citep{bath32820,wells-book}.
FRODA takes as input a decomposition of the protein into rigid and
flexible clusters provided by the FIRST (Floppy Inclusions and Rigid
Substructure Topography) software.
FIRST identifies geometric constraints related to bond lengths and
angles, hydrophobic interactions, and hydrogen bonds.
Using the pebble algorithm, these constraints are used to determine
flexible and rigid regions in the pigment-protein framework.
The rigid regions form so-called rigid clusters that, as we
shall show, display well-correlated motion due to their strong
intra-connectivity.
While the rigid clusters that form due to non-covalent interactions are the focus
of the study, it should be noted that the porphyrin ring of the
pigment molecules forms a  rigid cluster due to the structure of the covalent
bonds.
The central Mg atom of each pigment was rigidly connected to the
protein {\it via} the ligands identified in the X-ray crystal
structure.
Here, to generate rigid cluster decompositions, the hydrogen bond
cut-off energy (E$_{cut}$) was set to $-4.6$~kcal/mol.
The principal component subspaces spanned by FRODA simulations have
been shown to be very robust with respect to the chosen value of
E$_{cut}$ \citep{david2011}.

FRODA (Framework Rigidity Optimized Dynamic Algorithm) efficiently
simulates coarse-grained motion using the pre-determined geometric
constraints output by FIRST.
To generate a new conformation, the atom positions are perturbed in
random directions by a magnitude of 0.1~\AA{}.
This is followed by an iterative cycle that allows the constraints to
be efficiently and accurately re-introduced by fitting the atomic
positions back on to so-called ``ghost templates'' (which store the
information relating to the previously identified constraints).
It should be noted that, unlike MD, FRODA is not a dynamical technique and hence does not supply any characteristic time scales for the formation of individual structures.
By significantly reducing the allowed degrees of freedom, FRODA is
able to efficiently sample the relevant subspace of the total
conformational search space.
Such efficiency does come at the cost of being limited to a fixed
constraints topology, generating an athermal ensemble and neglecting
long-ranged electrostatic interactions and solvation (except for a
small number of explicitly included water molecules).
Yet even within these approximations, FRODA-generated atomic
fluctuations have been shown to agree with those identified in MD
simulations and, critically, NMR experiments \citep{FRODA_beats_MD}, indicating that
dynamics of the native state emerge naturally from a simple network of
contacts.

FRODA was employed, in this study, to generate an ensemble of allowed
structures of the FMO pigment-protein complex.
We have performed 16 independent simulation runs and each has
generated 80\,000 distinct conformations.
In order to gain a representation of the conformational ensemble,
every 250th structure was stored for further analysis.
Monomer 1 (as labeled in the PDB structure) was analysed in two
simulations; a simulation in the absence of the other two monomers
(the monomer simulation) and a simulation of the full trimeric
structure (the trimer simulation).
We have performed both monomeric and trimeric simulations, though all
of the analysis that follows is with respect to the trimer simulation
except where explicitly stated.

Within AMBER, the improved matrix facility of PTRAJ~\citep{matrix} was
employed to generate the root-mean-square fluctuations (RMSF) of the
$C_{\alpha}$ atoms of the stored conformations.
The cross-correlation coefficient between pairs of atoms was also
generated under this environment as follows:
\begin{equation}
C_{ij}=\frac{\left\langle \triangle r_{i}\cdot\triangle r_{j}\right\rangle }{\sqrt{\left\langle \triangle r_{i}\cdot\triangle r_{i}\right\rangle \left\langle \triangle r_{j}\cdot\triangle r_{j}\right\rangle }}
\end{equation}
A score between two residues can range from +1 (correlated) to -1
(anti-correlated).
This analysis allows the identification of long-range correlations
present in the conformational ensemble, which may not be obvious from
direct visualisation of the structure alone.
Principal component analysis (PCA) was employed to extract
biologically relevant motion using the Bio3d software package, which
employs the R environment and has powerful statistical computing
capabilities to generate the PCs~\citep{bio3d}.
The lowest frequency modes identify the large amplitude concerted motions and are, therefore, intimately related to protein function and important in elucidating how the FMO complex is
able to preserve quantum coherence in a noisy
environment~\citep{imp_low_Freq_mode}.

A fundamental process that occurs in PPCs, such as in the FMO complex,
is resonance energy transfer.
In a system of pigments with no orbital overlap, excitation energy
transfer is mediated by Coulombic coupling between the transition
densities of the chromophores.
In the FMO complex, the coupling interaction between pigments $m$ and
$n$ (V$_{mn}$) has been shown to be well-reproduced by the point
dipole approximation~\citep{pda}.
This approximation uses the strength and direction of the transition dipole moment to estimate the distribution of excitonic
couplings that are present in the ensemble of FRODA-generated structures according to: 
\begin{equation}
V_{mn}=f\frac{\mu_{vac}^{2}}{R_{mn}^{3}}[\vec{e}_{m}\cdot\vec{e}_{n}-3(\vec{e}_{m}\cdot\vec{e}_{mn})(\vec{e}_{n}\cdot\vec{e}_{mn})]
\label{eq:PDA}
\end{equation}
where $\vec{e}_{mn}$ is a unit vector joining the Mg ions of pigments
$m$ and $n$ and $\vec{e}_{m}$ is a unit vector along the transition
dipole moment of pigment $m$ (defined by the line joining the N atom
of pyrrole rings I and III).
$\mu_{vac}=6.1$~D is the transition dipole strength of the Bchl
pigment in vacuum and its value has been
experimentally-determined~\citep{knox03}.
The factor $f$ describes the enhancement of the transition dipole
moment in the protein environment, dielectric screening of the
Coulombic interactions, and neglect of the dielectric cavities of the
pigments.
It has been shown, in comparisons with full solutions of the
inhomogeneous Poisson equation, that $f=0.8$ is a suitable scaling
factor for the FMO complex.
A more detailed discussion of the point dipole approximation can be
found in~\citet{Adolphs:2006yq}.

\section{Results}
\label{sec:results}
The trimeric structure of the FMO complex comprises 19,863 atoms.
Therefore, to efficiently simulate motion on biologically relevant timescales we
must employ some approximations to reduce the degrees of freedom of present in
the system.
To this end, Figure~\ref{fig1}(c,d) shows a rigid cluster
decomposition of the FMO complex using the FIRST software at a
hydrogen bond energy cutoff ($H_{cut}$) of $-4.6$~kcal/mol.
An unusual feature of the complex is the persistence of a large rigid cluster at
low values of $H_{cut}$.
This rigid cluster encompasses the majority of the clam shell (CS) architecture, which
sequesters the optically-active pigments from the environment.
The implications of the size and robustness of the CS rigid cluster will
be discussed later.
The internal structure of the complex, which mostly
comprises the pigments and $\alpha$-helices, is by 
comparison relatively flexible. Recall, it is widely accepted that 
the $\alpha$-helices play a role in modulating the site energies of the pigments.
Our choice of $H_{cut} = -4.6$~kcal/mol, therefore, allows us to
efficiently explore the motion of the pigments and their environment within the
approximation of a rigid CS structure.
\begin{figure*}
\includegraphics[width=1.0\textwidth]{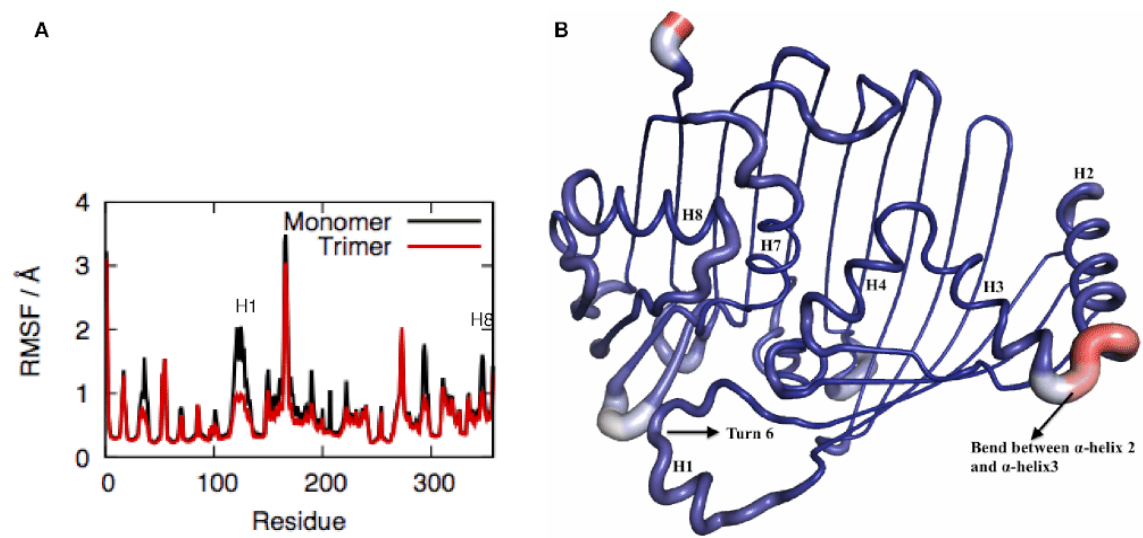}
\caption{(A) RMSF (\AA{}) per residue in monomer 1 from the monomer and
  the full trimer FRODA simulations. (B) The RMSF values from the trimer
  simulation have been displayed by colour (increasing
from blue to red) and tube thickness.} 
\label{fig2}
\end{figure*}
The FIRST rigid cluster analysis gives valuable information regarding the
expected flexibility in the FMO complex.
We have employed the FRODA software to further explore its allowed
conformational space and elucidate the structural dynamics supported
within these static results.
FRODA took, as an input, the FIRST rigidity analysis at $H_{cut} =
-4.6$~kcal/mol, which included 713 hydrogen bond and 236 hydrophobic
non-bonded constraints.
The RMSF calculated for each residue, averaged over the ensemble of
output configurations, has been displayed in Figure~\ref{fig2}.
In the trimer simulation, the analysis reveals few regions with high
conformational mobility -- most residue fluctuations are lower than
1~\AA{}.
One exception is the bend that connects $\alpha$-helices~2 and~3 (residues
close to 166). This is most likely due to the presence of three sequential glycine
residues, which support the greatest backbone mobility. Note $\alpha$-helices~2 and~3 
are found to form direct non-bonded interactions with Bchls~1 and~2.
A broad, mobile region is found from residues 120 to 130, which
includes $\alpha$-helix 1 and several residues either side.
Another broad peak has been identified that encompasses
$\alpha$-helices~2 through 4.
Small peaks that are caused by subtle motion of $\alpha$-helices~7
and~8 can also be seen.
Interestingly, the N- and C- terminal ends of $\alpha$-helix~8 and the
N-terminal end of $\alpha$-helix~7 are very close in sequence to
several small $\beta$-sheets (that do not belong to the CS
rigid cluster).
Furthermore, these $\beta$-sheets are inter--connected via several non-bonded
interactions and are more mobile due to their position on the end of
the CS.

In the monomer simulation, we found $\alpha$-helix~1 to have an RMSF
that is more than twice that found in the trimer simulation.
This is understandable because $\alpha$-helix~1 forms direct contact
with adjacent monomers in the full trimer structure.
Less expectedly, the monomer also demonstrated much larger motion of
$\alpha$-helix~8 and turn~12 than in the trimer, and the bend
between $\alpha$-helices~2 and~3 was also found to increase slightly.
Recall that $\alpha$-helices, in particular $\alpha$-helix~8, have
been shown to strongly influence the optical site energies of the
pigments~\citep{direct}.
We therefore emphasise that the full trimeric structure should be used
in dynamical studies of environmental fluctuations in the FMO complex.

Concerning the intramolecular atomic fluctuations of the pigments, we speculated that
the average RMSF for each pigment may be related to the site energy
disorder, and therefore the static inhomogeneous width and structure
of the excitonic lineshapes in the linear absorption and linear and
circular dichroism
spectra~\citep{van2000photosynthetic,Adolphs:2006yq,atom_study,cole2013}.
However, there is no significant difference in the RMSFs of pigments
1--7, averaged over the atoms of the porphyrin rings, which are small
and range from 0.36 (Bchl~6) to 0.77~\AA{} (Bchl~5).
The contrast between these weak fluctuations and the much larger
differences in site energy distributions identified by the MD
simulation in ~\citet{atom_study}, which generates a particularly
broad spectral function on site~7, suggests that the site energy
dispersion depends more strongly on interactions with the protein
environment.
The static influence of slow, conformational protein motion was not
accessible to these techniques, and we now consider these in more
detail.

We have already shown that the Bchl pigments of the FMO PPC, protected
by a remarkably robust CS structure, display low
conformational flexibility.
This is particularly evident when comparison is made with the rigidity
of structures that have previously been investigated with FRODA
analysis \citep{froda_compare}.
Following~\citet{froda_compare}, rigidity was
measured by counting the number of C$_{\alpha}$ atoms that belong to
the five strongest rigid clusters at $H_{cut} = -3.0$~kcal/mol in relation to the
total number of C$_{\alpha}$ atoms (denoted $f_5$, see Fig. 7 in
\citet{froda_compare}).
Of the 51 proteins analysed in \citet{froda_compare}, only three
proteins from a particular rigid protein family had a higher ratio
than that revealed here for the FMO complex ($f_5 = 0.56$).
Furthermore, the FMO complex is constructed from more than twice the
number of residues as the most robust protein structures studied in
~\citet{froda_compare}.

\begin{figure}
\includegraphics[trim = 40mm 25mm 45mm 5mm, clip, width=0.45\textwidth]{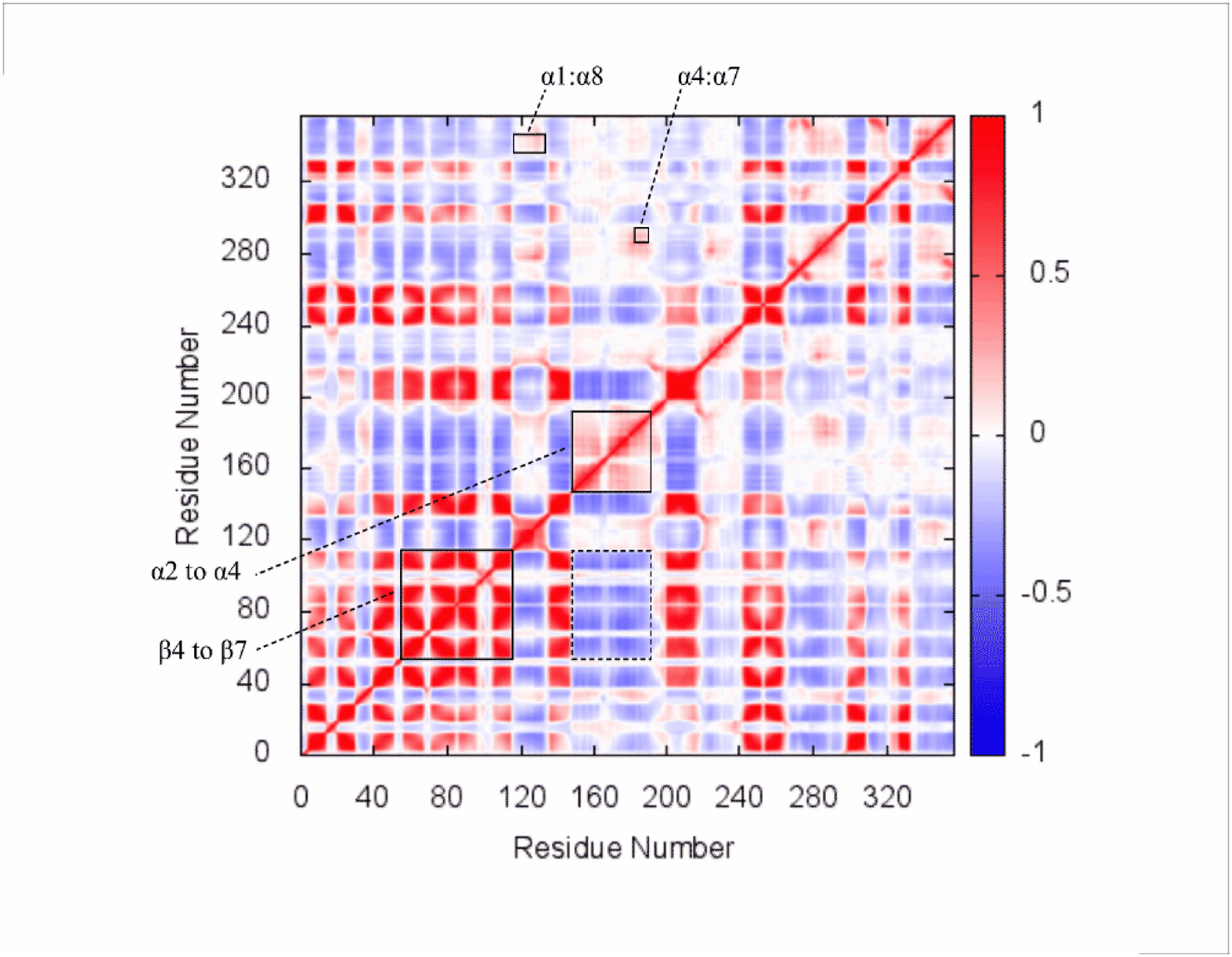}
\caption{The cross-correlation matrix for monomer 1 in the trimer
  simulation. Several noteworthy correlations between $\alpha$-helices
  are highlighted. Also shown are $\beta$-sheets 4 through 7, which
  make up a large proportion of the CS structure -- note also,
  that these structures are anti-correlated with the
  $\alpha$-helices.}
\label{fig3}
\end{figure}
A minimisation of flexibility (and thus thermal exploration of
conformational states) of key structural elements (the CS and
pigments) acts as a primary mechanism in the
protection against the disorder of the Hamiltonian parameters which could
disrupt the basic structure of the EET pathways. This
 is an expected requirement given the function of this pigment-protein complex.
However, the seemingly excessive rigidity of the FMO protein suggests
that this super structure may have been selected for purposes
additional to simply holding the pigments in position. Indeed, we now show how this rigidity promotes and coordinates substantial
correlations in the more flexible secondary elements of the structure.
Computing the average cross-correlations between residue pairs in the
ensemble of output structures from the FRODA simulation
(Figure~\ref{fig3}), we see that the residues belonging
to the CS rigid cluster are extremely well-correlated over large distances.
Interestingly, these residues are also, generally, anti-correlated
with the functional $\alpha$-helices in the protein.
It was also found that $\alpha$-helices~4 and~7, as well as
$\alpha$-helices 2--4, displayed correlated motion with each other.
The correlation between $\alpha$-helices 2, 3, and 4 is particularly
noteworthy given that they span a distance of roughly 24~\AA{} and
implies that long-range communication is present in the protein
environment.
$\beta$-strands 17, 19, and 20 are found to be flexible and reside at
the end of the CS structure near $\alpha$-helix~8.
These strands are also anti-correlated with the CS rigid cluster, and
display some correlation with the $\alpha$-helices.
This provides further evidence that the CS rigid cluster is involved in
the maintenance of globally correlated conformal movement.
%
\begin{figure*}
\includegraphics[width=1\textwidth]{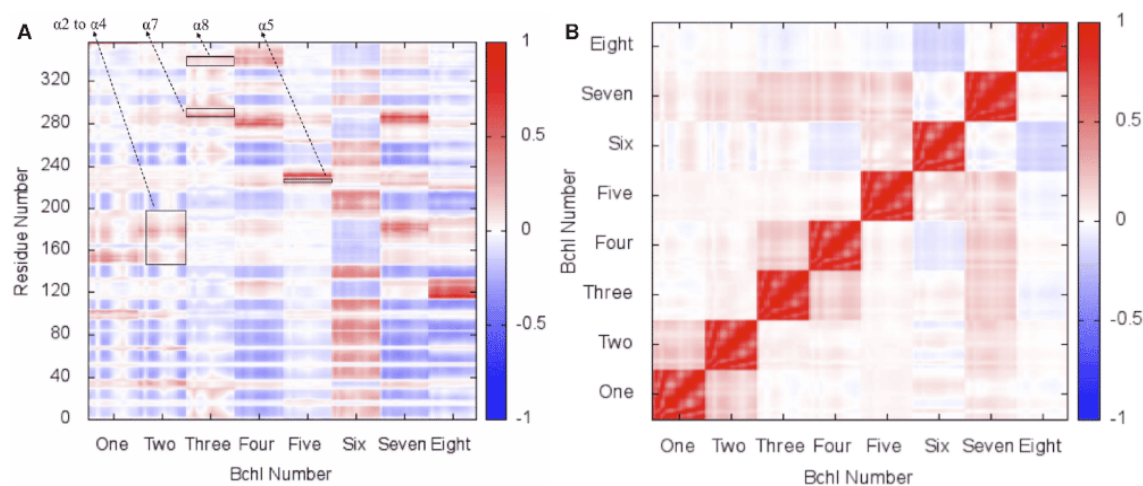}
\caption{(A) Correlations between the pigments and important
  secondary structure elements of the protein. (B) Correlations between the 8 Bchl pigments of the FMO
  complex.} 
\label{fig4}
\end{figure*}
Our results also demonstrate evidence for conformational correlations
between pigments that are theoretically predicted to display strong
excitonic coupling and form delocalised states
\citep{Adolphs:2006yq,Adolphs:2008fu} namely; between Bchl 1 and 2, 
Bchl 3 and 4, Bchl 5 and 6, and Bchl 7
with Bchl 1--5 (Figure~\ref{fig4}(B)).
Figure~\ref{fig4}(A) shows the presence of correlations between the
pigments and the protein environment.
Firstly, with the exception of Bchl~6, the pigments display
anti-correlated motion with the CS RC.
Moreover, important correlations have been identified between secondary
structure elements in the protein that, critically, modulate the
pigment site energies.
Bchls 1, 2, and 7 display correlated motion with $\alpha$-helices 2
through 4.
It is likely that this correlation gives rise to the observed
correlated motion with $\alpha$-helix~7.
Bchls 3, 4, and 7 also show strongly correlated motion with
$\alpha$-helices 7 and~8 (or residues adjacent in sequence).

\begin{figure}
\includegraphics[width=0.45\textwidth]{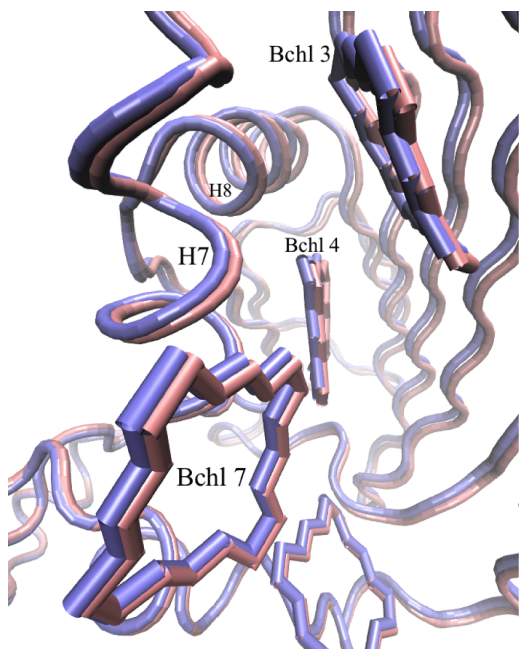}
\caption{The synchronised motion of $\alpha$-helix 7 with pigments 3,
  4, and 7 that is exemplified by the lowest frequency principal
  component.}
\label{fig6}
\end{figure}
Principal component analysis of the ensemble of structures from the
FRODA simulation reveals the low frequency modes of the system, and
allows us to visualise large-amplitude, concerted motions that may be
relevant at long time scales.
Here, the lowest frequency principal component provides additional
evidence for the concerted movement of the functional $\alpha$-helix~7
with pigments 3, 4 and 7 (Figure~\ref{fig6}).
%
%
The PCA has also identified low frequency motion for $\alpha$-helices
1--4 and 8.
%
%
%
%
It has been shown previously that the electrostatic dipoles of $\alpha$-helices
can significantly alter the local site energies of
pigments~\citep{direct,Adolphs:2008fu,cole2013}.
In particular, $\alpha$-helices~7 and~8 are thought to be major
contributors to the energy sink formed around pigments~3 and~4.
Our observations of correlated (Figure~\ref{fig4}(A)) and concerted
(Figure~\ref{fig6}) motion between pigments~3 and~4, and
$\alpha$-helices~7 and~8, make it highly plausible that spatial site
energy correlations also exist.
Such a mechanism would preserve the relative site energies of the pigments in the configurational ensemble and would potentially maintain Bchl 3 as the ultimate sink for EET in each conformational realisation of the complex. However, as mentioned previously, uncorrelated dynamical fluctuations on EET timescales (ps) are still required to drive fast energy transfer in the static energy landscape enforced by these correlated structures \citep{NMA_protein, correlation, QC_role}. 

\begin{table}
\caption{The inter-pigment excitonic couplings in trimer simulations
  of the FMO complex (cm$^{-1}$). The couplings identified by \citet{Adolphs:2006yq}
 (column 2) are compared with the
  mean excitonic couplings identified in the current analysis (column
  3). The standard deviation for these couplings over the ensemble of
  output structures is displayed in column 4, where the coupled
  pigments displaying large $\sigma$ have been highlighted in bold.}
\label{tab1}
\begin{tabular}{cccc}
\hline\noalign{\smallskip}
Pigments & Ref.~\citet{Adolphs:2006yq} & This Work & $\sigma$ \\
\noalign{\smallskip}\hline\noalign{\smallskip}
1-2 & -98.2 & -92.8 & 7.8\\
1-3 & 5.4 & 5.8 & 0.7\\
1-4 & -5.9 & -5.9 & 0.6\\
1-5 & 7.1 & 7.1 & 0.9\\
1-6 & -15.2 & -14.6 & 4.5\\
1-7 & -13.5 & -9.8 & 2.0\\
2-3 & 30.5 & 29.4 & 2.0\\
2-4 & 7.9 & 6.4 & 0.8\\
2-5 & 1.4 & 1.5 & 1.2\\
2-6 & 13.1 & 10.9 & 1.2\\
2-7 & 8.5 & 2.7 & 2.9\\
\textbf{3-4} & -55.7 & -47.0 & 13.0\\
3-5 & -1.8 & -0.1 & 1.5\\
3-6 & -9.5 & -9.4 & 0.4\\
\textbf{3-7} & 3.1 & 15.8 & 10.0\\
4-5 & -65.7 & -58.8 & 7.3\\
4-6 & -18.2 & -16.1 & 1.5\\
4-7 & -58.2 & -63.1 & 7.4\\
\textbf{5-6} & 88.9 & 91.3 & 14.0\\
5-7 & -3.4 & -2.6 & 5.4\\
6-7 & 36.5 & 35.0 & 5.7\\
\noalign{\smallskip}\hline
\end{tabular}
\end{table}
The formation of excitonic delocalised states is dependent on the
extent of the dipole-dipole coupling interactions between the pigment
molecules, where strong excitonic coupling gives rise to collective electronic
excitations that have a high degree of delocalisation.
This phenomenon depends strongly on the transition dipole moment which
underlies the dipole-dipole coupling interaction V$_{nm}$ between two
pigments, which is, in principle, a sensitive function of the protein
structure through its dependence on pigment-pigment separation and
orientation.
As discussed previously, an approximation of the coupling strength
can be estimated from the transition dipole moments obtained from
purely structure-based methods \citep{cole2013,Adolphs:2006yq,direct}. 
Importantly, this analysis does not require costly quantum mechanical
 calculations.
Table \ref{tab1} shows the excitonic couplings and their standard
deviations $\sigma$ for monomer 1 from the trimer
simulations. These values are compared with a previous
 study that used an identical method but was based only on the static crystal structure~\citep{Adolphs:2006yq}.
The mean excitonic couplings of the ensemble of FRODA-generated
structures, in general, agree with the calculations performed by
~\citet{Adolphs:2006yq}.
%
%
However, there are several noteworthy deviations between the
previously calculated static couplings and the present data,
which includes the effects of environmental fluctuations 
through the inclusion of structural  flexibility. The largest deviation was found to be 
that between pigments~3 and~7.
While the coupling calculated directly from the crystal structure was
low (3.1~cm$^{-1}$), the current analysis has found that the
averaged coupling over the conformational ensemble is higher, albeit
with relatively large fluctuations ($15.8\pm 10.0$~cm$^{-1}$).
%
%
A similar situation is identified for Bchls~3 and~4, as well as Bchls~5 and~6.
The coupling between Bchls~3 and~4 displays the
second largest deviation from the static coupling and also second
highest $\sigma$.
This link is also a part of the major EET pathway proposed for
excitations entering the FMO complex via site 6
\citep{Adolphs:2006yq,mohseni2008environment,plenio2008dephasing,caruso2009highly}.
\begin{figure}
\includegraphics[width=1\textwidth]{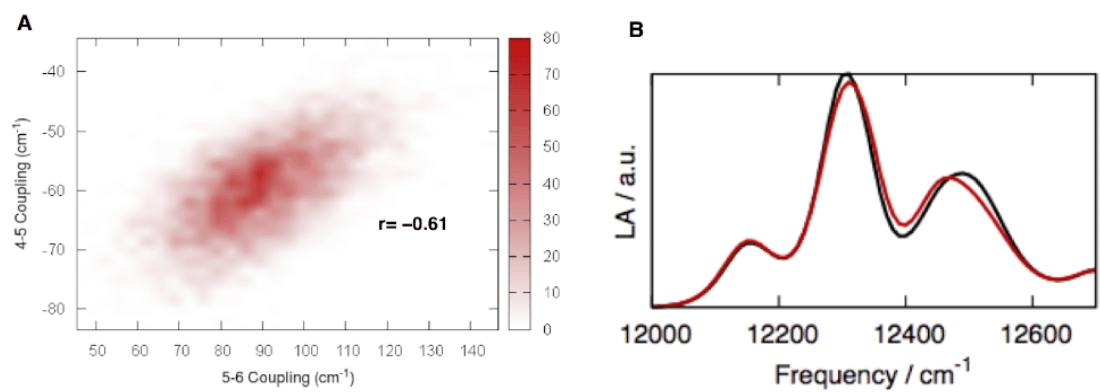}
\caption{ (A) Heat map displaying correlations between the excitonic
couplings found between Bchls~5 and~6 and Bchls~4 and~5. The negative Pearsons
coefficient suggests that as the magnitude of coupling between Bchls~5 and~6
strengthens, the coupling between Bchls~4 and~5 diminishes. (B) Simulated linear 
absorption spectra for FMO using parameters described in the text. The plots 
show the spectrum computed using
the mean values of the excitonic couplings (red line) and by
averaging over the distribution of coupling strengths found by our
FRODA-based analysis (black line). The high rigidity and
conformational correlation between structures leads to very low
spreads of couplings and negligible variation in the spectrum.}
\label{coupling_correl}
\end{figure}
The small $\sigma$ identified between Bchls~2 and~3 (2.7 cm$^{-1}$) is
a significant finding. This  shows that the excitonic coupling in an
important link in one of the main EET pathways in the FMO
complex~\citep{Adolphs:2006yq,mohseni2008environment,plenio2008dephasing,caruso2009highly,chin2010noise,caruso2010entanglement,rey2013exploiting}
is preserved despite thermal motion in the environment.
The robust conservation of the coupling between these two pigments can
be attributed to the conservation of their transition dipole
orientation.
This is predicted to be accomplished by a bend between
$\beta$-strands~2 and~3.
This bend is (a) able to span the distance between the two pigments
due to its long length and (b) has the highest concentration of
prolines in the amino acid sequence.
Recall that proline residues can be employed to rigidify loops
\citep{proline_rigidity} due to their ability to reduce the
conformational flexibility of the protein backbone.
It may be that a conformationally restricted bend is a ``just right''
measure that allows the two pigments to remain connected while also
allowing communication with other functional elements
($\alpha$-helices, pigments, etc.).
Although the principle findings with relation to the excitonic coupling
is their small variability, correlations in this electronic property
were also discovered. For example, Figure~\ref{coupling_correl}(A) shows the correlations in the excitonic
 coupling fluctuations identified between Bchls~5 and~6 and Bchls~4 and~5. Several other
correlations were also found between Bchls~3--4 and~4--6 (R=+0.62) and between Bchls~4--7 and~3--7 (R=-0.43).

Finally, to investigate whether the excitonic structure of the FMO
complex is robust with respect to the, apparently small, coupling
fluctuations shown in Table~\ref{tab1}, we have computed their impact
on the linear absorption spectrum of the PPC at $T=77$~K.
The red line in Figure~\ref{coupling_correl} is computed using the trimer
pigment site energies taken from ~\citet{Adolphs:2006yq} and the
mean excitonic couplings from Table~\ref{tab1}.
The theoretical treatment of the spectrum (Hamiltonian and secular
Redfield theory), as well as all other parameters (spectral function, inhomogeneous disorder) are as described in ~\citet{cole2013}. As can be seen in Figure~\ref{coupling_correl}, averaging the spectra over the distribution of coupling strengths, which are assumed to be independent, leads to negligible difference in the optical spectrum (black line). The spectra are instead dominated by homogeneous and site energy disorder. 
If our hypothesis that the strongly correlated spatial structures of
each realisation of the complex also lead to correlated site energies
in each ensemble member, then the excitonic structure and EET pathways
will be highly conserved across the conformal ensemble. Such a mechanism for robust EET pathways and performance could be important for organisms such as GSB which make use of many thousands of FMO complexes in their functional photosynthetic apparatus.
\section{Discussion}
\label{sec:discussion}
\subsection{The Role of Motion in Protecting Efficient Energy Transport}
The existence of perisistant coherence in experimental ensemble signals relies heavily on suppressed disorder of the site energies of the coupled pigments in each complex and also between members of the experimental ensembles. 
As discussed above, these energies are significantly modified by the
surrounding environment. In the current study, we present evidence for the
(natural) selection of correlated conformal motion in the 
FMO complex. In light of previous work \citep{QC_role}, 
we speculate that this behaviour facilitates a means for
 correlating fluctuations in pigment
 site energies. This
would allow
realisations of the FMO protein that have significantly different band gaps
between
their manifold of optical excited states and the ground states to preserve
the all-important energy differences between the states in the optically
 excitable manifold \citep{lee2007coherence,fidler2011two}.
This design principle would naturally minimise the population of
 conformations that worsen or halt energy transfer
(for example, by creating new energetic minima).
By employing key structural correlations these
disfunctional conformations are likely to have been selected against
during the evolution of GSB. 
However, we draw the readers' attention to the fact that the
significant correlations of atomic motion that we observe do not
guarantee correlations of all electronic properties of the pigment
network. While the impact of low frequency conformal changes in cases
where electronic properties depend almost trivially on spatial
coordination, such as dipolar excitonic coupling, is immediately
clear, the effect on local site energies is much more complex. For
instance, a recent MD study of \citet{correlation} found correlations of
fast atomic motion that did not result in significant correlations of
local site energy fluctuations \citep{correlation}. Similarly, \citet{NMA_protein} found
correlations of atomic motion via a normal mode analysis that resulted in
a negligible impact on dynamical relaxation and decoherence.  To
fully establish the link between site energies and atomic and rigid
cluster displacements will require further structure-based
calculations on the wide range of configurations generated by our
sampling methods, using techniques such as those recently reviewed by
\citet{renger13}. In spite of this important caveat, the
computational efficiency of the FRODA technique allows us to
potentially explore relatively large and correlated motions that, as stated previously,
go beyond the harmonic approximation. We expect this property would lead to
some degree of correlation in the inhomogeneous distribution of
excitonic energies in an ensemble of FMO complexes that were not
evident in \citet{correlation} and \citet{NMA_protein}. This is suggested by the significant correlated
motion of pigments relative to the $\alpha$--helices whose dipolar
electric fields often dominate the local site energy shifts on each
BChl. Other arguments for how these conformal motions may lead to
correlated energy shifts are highlighted in the following detailed
discussions of the hierarchy of motions that we report for FMO, though
direct confirmation by further work is still required. We now discuss 
the details of these conformal structural relations and their correlations in more detail.  

\begin{figure}
\includegraphics[width=0.45\textwidth]{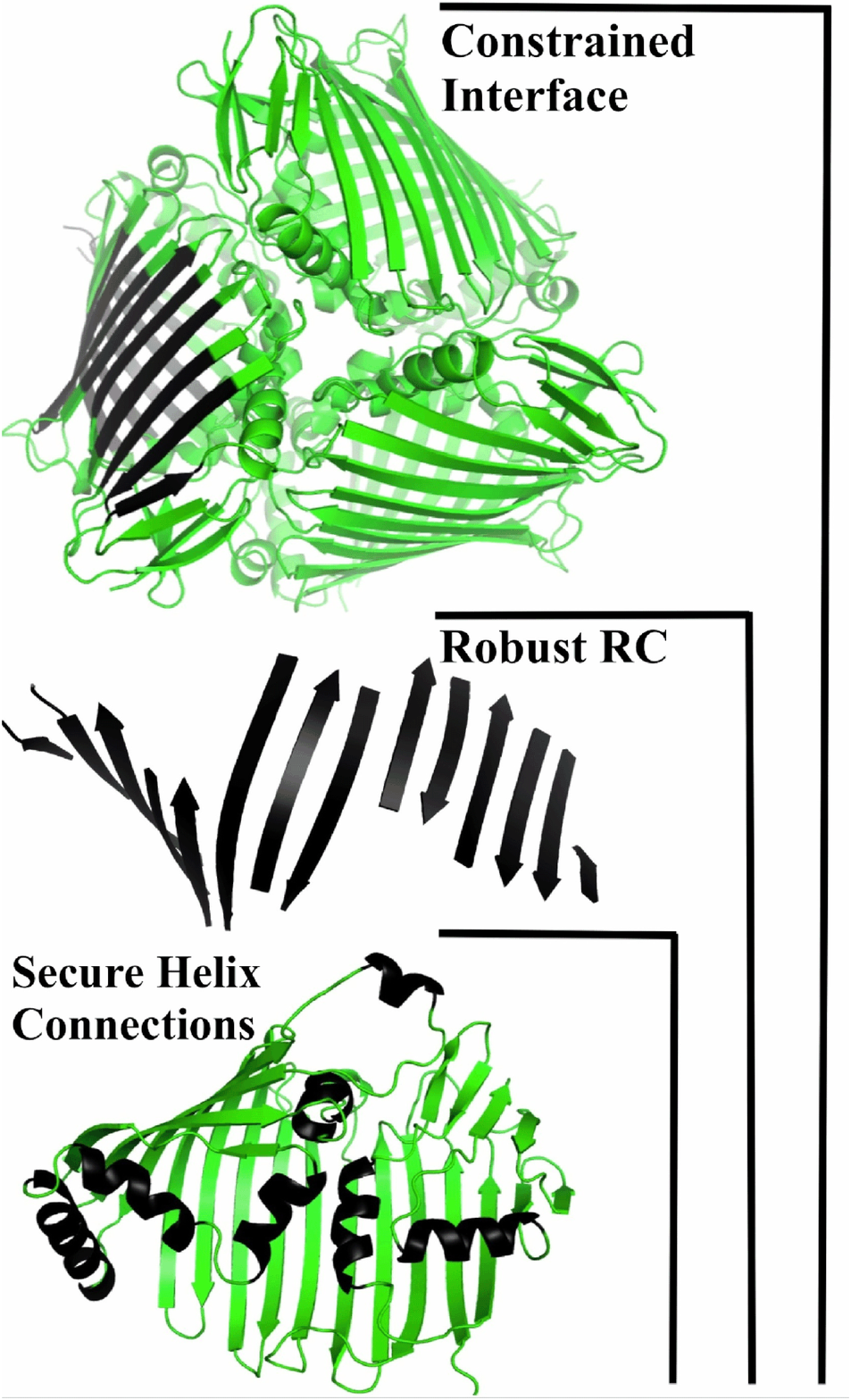}
\caption{Non-covalent interactions are employed to reduce the conformational fluctuations (uncertainty)
in the motion of the FMO complex. As displayed above, these interactions
can be grouped together into three main bands. Importantly, the observed correlations
 allow communication between the groups, an effect we have termed
\it{trickle down structural organisation}.}
\label{fig7}
\end{figure}

We hypothesise that the FMO PPC uses three main levels of structural
organisation to reduce mobility and constrain the
conformations of the protein to ones that support
coherent EET (Figure~\ref{fig7}).
In what follows, we discuss the mechanisms by which specific non-bonded
interactions preserve correlated motion between functional elements of
the PPC in each of these three layers of structural organisation, thus 
safeguarding efficient EET.
\subsection{Hydrophobic Interactions are Important for Inter-Pigment Correlations}
\begin{figure}
\includegraphics[width=1\textwidth]{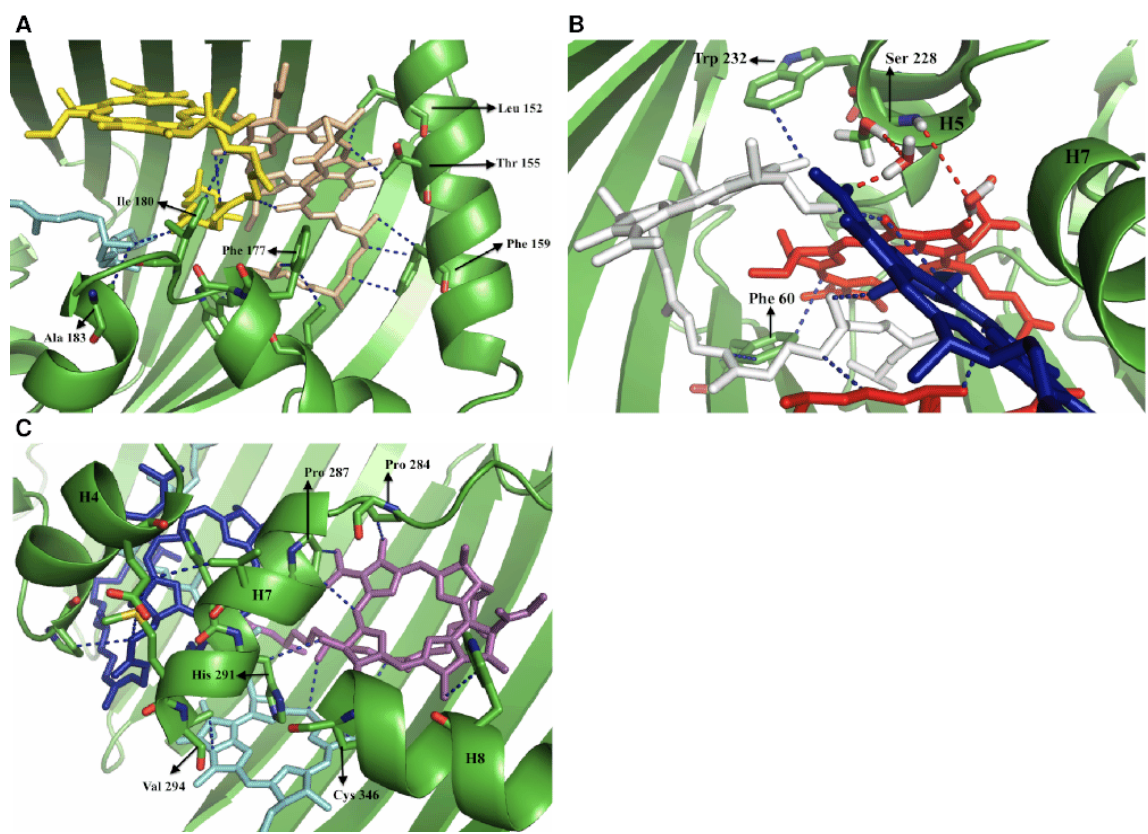}
\caption{(A) Extensive hydrophobic interactions between Bchl 1 (wheat) and $\alpha$--helix 2 can be seen.
 Several hydrophobic interactions between Bchl 1 and Bchl 2 (yellow) are shown.
 Hydrophobic interactions are also found with residues in (Phe 177) and adjacent to (Ile 180) 
$\alpha$--helix 3. 
(B) A hydrophobic interaction was identified between Bchl 6 (white) and turn nine residue Trp 232.
 The hydrophobic cluster that forms between Bchl 5 (red), 6, and 7 (blue) has been shown. 
A hydrogen bonding cluster was also identified, which involved Bchl 5, Bchl 7, Ser 228, and two
water molecules.
 (C) Hydrophobic interactions are found to occur between Bchl 3 (cyan) and Bchl 4
 (magenta). Bchl 3 and 4 are also found to interact with $\alpha$--helix 7 (via Val 294 and His 291,
 respectively) and $\alpha$--helix 8 (via Cys 346). Hydrophobic interactions were also identified between 
 Bchl 4 and residues near in sequence or in $\alpha$--helix seven. Note, an interaction was also identified with
with  His 291, which coordinates Bchl 3. A hydrophobic cluster involving Bchl 7, $\alpha$--helix
 4, and $\alpha$-helix 7 can also be seen.}
\label{fig8abc}
\end{figure}
EET is initiated from either Bchl~1 or~6, after which excitons relax
towards Bchl~3 for entry to the RC \citep{exp_EET}.
The nearly degenerate Bchl~1 and~2 have been predicted to form a
strongly delocalised pair of excitonic states, leading to a
delocalisation-induced energy splitting and enhancement of EET in the
pathway from sites~1 to~3~\citep{chin2010noise}.
Indeed, it has also been proposed that the excitonic splitting might
be tuned to exploit a specific frequency of the environmental noise to
achieve this enhancement, as a result of the so-called `phonon
antenna' mechanism~\citep{rey2013exploiting}.
These quantum EET pathway dynamics require that the energy differences
between sites 1 and 2 remain smaller than their excitonic coupling.
The current analysis shows that these pigments display correlated motion, and that
their excitonic coupling varies negligibly across different
conformational realisations.
Although Bchl 1 and 2 interact directly with each other, their
correlated movement is also maintained via interaction with nearby
$\alpha$-helices that display correlated motion with the pigments
(namely, $\alpha$-helices 2 through 4, see Figure \ref{fig8abc}(A)).
Bchl 5 and 6 display correlated motion, and have been shown
experimentally to facilitate coherent EET \citep{exp_EET}.
This is accomplished through non-covalent
interactions between the two pigments (Figure \ref{fig8abc}(B)).
Trp 232 is also well positioned to facilitate an indirect contact
between Bchl 6 and turn 9 (which itself coordinates Bchl 5).
The identified correlation between Bchl 5 and 7 relies on interactions
with overlapping biological elements.
Bchl 5 displays correlated motion with $\alpha$-helix 5, which
precedes turn 9 and, in turn, interacts with $\alpha$-helix 7 and
thereby provides an indirect connection between Bchl 5 and 7. These
correlations are partially mediated by hydrogen bonds with water molecules.

It has been found that all the pigments (except Bchl 6, which will be
discussed in the next section) are correlated with Bchl 7.
From Bchls 5 and 6 the EET relaxes to Bchls 4 and 7. These two pigments
are also found experimentally to participate in the same EET pathways
\citep{exp_EET}.
Indeed, Bchl 3 and 4 interact directly with Bchl 7. However,
correlated motion is presumably enhanced further  by their mutual interactions
with $\alpha$-helices 7 and 8 (Figure \ref{fig8abc}(C)). Critically, these $\alpha$-helices have previously
  been shown to
affect the site energies of these pigments \citep{Adolphs:2008fu}.
The sensitivity of Bchl 7 to the motion of the other pigments suggests
that it may play a unique role in the EET cascade.
The importance of Bchl 7 has already been noted by~\citet{subsystem}
 in connection with their research of quantum 
redundancy in the FMO complex.
Therefore, although Bchl 7 is only present in one pathway, it may be a vital component for conserving important correlations in both EET pathways. This idea is further supported by the striking demonstration of \citet{Bchl_7_imp}. Namely, optimised energy transfer through the FMO complex, as a function of temperature, is achieved by \emph{maximal} excitonic correlations between site 7 and the rest of the pigments of the FMO complex.
The correlations suggest that the pigment conformal motions remain in
unison through \emph{slow} correlated fluctuations managed by secondary structure
elements. We anticipate that correlated motions translate into
spatially coordinated changes in the energy landscape that, on the timescale of EET, provide static maintenance of the key EET pathways.

\subsection{The Robust CS RC}

The user chosen $H_{cut}$ was $-4.6$~kcal/mol. This allowed investigation of a
network of interactions that (a) are stable (due to the extremely low
cutoff) and therefore have the largest influence on the {\it in vivo}
energy landscape and (b) give rise to geometric structures in the
protein that are robust against time. 

Connecting regions of a protein with particular roles assigned by Nature can reveal subtle design principles that are employed to improve a particular function. The uniqueness of the CS RC is a significant observation, and implies that it plays a critical role. In fact, such a large RC encompassing several secondary structure elements at low values of $H_{cut}$ has not been seen in other systems \citep{froda_compare}. Indeed, the CS has an important structural role in the protein by sequestering the pigments and aiding the organization of the LHC by facilitating contacts with the baseplate, and perhaps promoting trimer aggregation in the periplasmic space between baseplates and RCs, which is another possible level of organisation which we do not deal with here. For example, the observation that 150-200 trimers contact the baseplate may demand tight packing that is facilitated by the CS RC \citep{blankenship}. However, the extrodinarily robust RC is also efficient at transmitting structural information over distances of more than 50~\AA{}.
In a previous work, the $\alpha$-helices were found to have an unanticipated ability to modulate the site energies of all seven sequestered pigments \citep{direct}. For example, the finding that Bchl 3 has the lowest site energy is a result of the dipoles of $\alpha$-helices 7 and 8, which are able to shift the site energy by around $-300$~cm$^{-1}$. Natural selection facilitates the accumulation of residues that improve a protein's function (in this case, energy transfer). Therefore, it can be assumed that Nature has employed $\alpha$-helices to manage the optical properties of the pigments and create an energetic funnel toward Bchl 3. The remarkable anti-correlation between the $\alpha$-helices and the CS suggest the existence of a higher level of communication, which further encourages correlation between the spatially separated functional units. Since the CS is well-correlated over large distances, it is an extremely well fitted candidate for mediating the long-range correlation of the functional $\alpha$-helices.
$\alpha$-helix 7 is strongly anti-correlated with the CS RC and dictates the movement of Bchl 7. Recall, this pigment is correlated with many of the pigments in the system. $\alpha$-helix 7 has bends either side that connect to the bottom and top of the CS. The reaction to the CS motion is therefore particularly strong. As a result, the motion of Bchl 6 (which is the only pigment that is correlated with the CS) seems to be anticorrelated with Bchls 4 and 7. The flow of information from the CS to $\alpha$-helix~7, and on to Bchl~7 and the other pigments is a possible approach employed by the FMO complex to reduce uncertainty and maintain a robust electronic landscape for EET. 
\subsection{The Constrained Monomer Interface} 
Relative to the monomer simulation, the trimer displayed a decreased
RMSF in several key regions of the FMO complex (Figure~\ref{fig2}).
One noteworthy example is the significantly decreased RMSF of
$\alpha$-helix 1 that occurs due to interactions with the adjacent
monomer.
It is important to note that $\alpha$-helix 1 is (a) strongly
anticorrelated with the motion of the CS in its own monomer and (b)
strongly correlated with the motion of the $\alpha$-helices and
pigments in the adjacent monomer.
Indeed, the motion of $\alpha$-helix 1 of monomer~1 is found to be
correlated with $\alpha$-helices 2 through 4 ($C_{ij}\approx0.5$), and
Bchls 1 and 2 ($C_{ij}\approx0.6$), of monomer~2.
As the FMO complex displays rotational symmetry, $\alpha$-helix~1 may
act to communicate between the functional units of the trimer and
thereby help constrain the conformations to (coherent) EET-competent
ones.
With the above observations in mind, it is predicted that the trimeric
structure adds a further level of organization that is exemplified by
the interactions of $\alpha$-helix~1.
\subsection{Trimer unit as a structure for maintaining energy throughput rates in the presence of functional failure}

\begin{figure}
\includegraphics[width=1\textwidth]{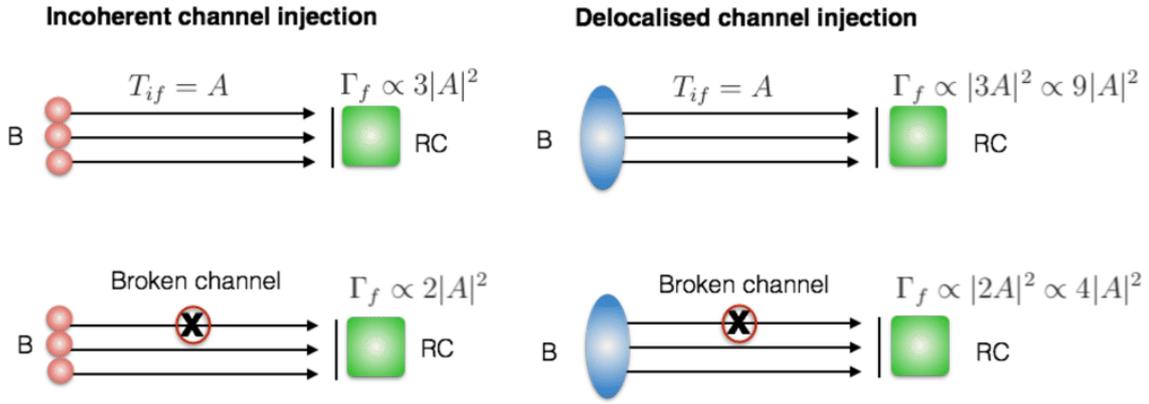}
\caption{A trimeric structure viewed as an ensemble of either classical or quantum wires. Quantum interference in the latter case can greatly increase average EET speeds over the classical wires, even in the case of a broken channel. }
\label{fig15}
\end{figure}
As a further and final speculation on the relevance of strong correlations in PPCs, we note that the use of a trimer structure -- aside from considerations of supra-molecular stability -- may also convey a degree of fault tolerance against failure of any of the individual monomers of the trimer, which could potentially be enhanced by quantum effects. Consider the situations depicted in Figure~\ref{fig15}, which shows an abstracted picture of the process of EET through the FMO to the RC. Following the notion of the FMO monomers as molecular wires, we associate with each line (representing a monomer) a transmission probability for propagating an exciton to the RC. In a classical kinetics, and assuming that each monomer has, on average, the same transmission probability, is independent of the others and is equally likely to receive the excitation from the baseplate, a broken channel (due to disorder, mutation, etc.) would reduce the overall transmission probability for the trimer by $1/3$ (Figure~\ref{fig15}). Of course if the transmission were effected by a single monomer, a failure would be total: the trimeric structure therefore generally helps protect against this (not in itself a new observation). Under conditions of continuous reception of excitations a failure would lead to an average rate of excitation delivery of $2/3$ of the fully functional rate. 

In contrast, many of the results we have obtained suggest that intra-- and inter--monomer disorder is suppressed to the extent that one could at least imagine the situation depicted in Fig.~\ref{fig15} (right). This shows a quantum analogue of the transmission process, where now the transmission is described in terms of the quantum mechanical amplitudes (propagators) $A$ for excitons to reach the RCs through each monomer. In this description, the initial condition plays a key role in the subsequent EET. If the excitation starts randomly (incoherently) on any of the monomers then the average transmission probability is the same as the classical case if we associate the square of the amplitudes to the classical transmission probability. However, if the initial condition is a coherent superposition over the three initial entry points to each monomer, the final amplitude of transmission (assuming each monomer couples to the same RC) would coherently sum to $A_{total}=\sqrt{3}A$, leading to a three-fold increase in transmission probability, $P=3|A|^2$ (meaning the rate of energy throughput is on average three times faster). This is a well-known (constructive) interference effect for waves, and appears in a variety of quantum contexts involving systems of common grounds states (such as absorption and emission of photons in coupled chromophores). If, in the quantum case, a channel is broken, the total amplitude is now $2A/\sqrt{3}$ leading to a probability of $P=4/3|A|^2$. Comparing with the classical case ($P_{cl}=|A|^2$), we see that the partially broken quantum wire still propagates excitations to the RC faster than even the undamaged classical transmission system. Although the quantum case maintains a higher propagation speed and average energy throughput, we note that in both cases a failure of a channel results in an overall loss of $1/3$ of the total number of excitations. 

This last idea is -- emphatically -- not a mechanism that we believe to operate in the FMO complex {\it in vivo}: the homogeneous dephasing rates between monomers is likely to be too great, and the injection mechanism from the baseplate may be too slow and random to prepare coherent superpositions of initial states. Indeed, it is not even obvious that enhanced transmission rates for excitons is something that is necessary or biologically desirable, although the idea that this structure -- or a modified version -- could adjust between trimeric and independent monomer operation, perhaps in response to metabolic needs or overload (to slow or speed up transfer), is an interesting idea for artificial quantum devices. However, these ideas are simply inspired from our observations of the extent to which the protein correlates intra-- and inter--monomer structures, and the (extreme) possibility that this could reduce decoherence across the trimer to facilitate injection into an initially delocalised state. While it is unlikely that this inter--monomer coherence could be sustained during the entire EET process, some recent theoretical work has shown that the first $50\%$ of an excitation injected into site $1$ of a monomer can propagate within a picosecond to the RC under Hamiltonian (coherent) dynamics alone, and -- as this fast part of the transfer retains its wave-like character in the presence of noise -- such pathways could be interfered \citep{mohseni2008environment,plenio2008dephasing,caruso2009highly,chin2010noise,caruso2010entanglement,rey2013exploiting}.      

\subsection{Conclusion and prospects}

In conclusion, we would like to suggest a number of ideas inspired by the wealth of structural information that we have gathered. We have stressed that the structural rigidity, correlated conformal fluctuations and inter--monomer interactions appear to suggest that the internal (monomer and trimer) energetic structure of the excited states may be conserved across different realisations of the protein and have shown explicitly that excitonic couplings show strikingly low variance across the ensemble of structures. This is particularly important for explaining the observation of quantum beats in 2D optical experiments, as the measured inhomogeneity of the one-quantum coherences would be expected to completely wipe out the zero-quantum (inter-exciton) coherences within 10-100 fs if the same disorder acted independently on each excited state. This is not what is observed. Preservation of the relative energy differences of the excited states, or signals dominated by sub-ensembles of particularly `coherent' conformations seem to be the only plausible explanation~\citep{lee2007coherence,fidler2011two,chin2013role}. Our results support the former picture, although both ideas are compatible with each other and are likely to contribute. We note that correlated disorder has recently been observed in the FMO complex~\citep{harel2012quantum}, although other novel microscopic mechanisms appear to be required to reproduce the homogeneous decoherence times observed for zero quantum coherences in FMO~\citep{christensson2012origin,kreisbeck2012long,chin2013role}.
Another possible explanation for the reduced effect of inhomogeneity has been suggested by \citet{christensson2012origin}, who suggest that the quantum beats seen in 2D experiments are actually of intra-molecular vibronic character and only suffer effects of (much weaker) inhomogeneity of the vibrational frequencies. Clearly, a substantial amount of further work which could profitably combine the present techniques and those which address the faster molecular motions are required to describe the rich, multilayer photophysics of this and other PPCs. 
Anticipating these studies, we speculate here that the robustness in electronic parameters is not achieved by simply having an extremely rigid structure, but also by extensive correlation of conformal motion. As is briefly mentioned below, it could be that this motion has been selected for as being fit for purpose, and we speculate that adapting or modulating \emph{motion or correlations} may be a much more energetically efficient and rapid way of responding to internal (external) needs (influences) than costly and possibly hard to reverse structural changes.

On a more general level, we believe that this study illustrates the
potential of the computationally very efficient rigidity analysis and
constrained geometric dynamics to tackle biomolecular problems in
which the identification of key functional or structural units (rigid
clusters), and a detailed, mechanistic description of their
interplay, cannot be determined from the static crystal structure
alone.
The present problem of FMO provides a good example of how these techniques can move us further toward the oft-stated goal of ``quantum biology'' to provide design principles for future light-harvesting devices and quantum technologies, doing so at the all-important level of supra molecular structure and motion. In this context, an exciting prospect is to consider mutated and altered structures within the framework here established, looking for structures with stronger conformal correlations, reduced overall motion of key residues and/or rigid clusters, or increased intermonomer interaction. Moreover, the feature of being able to access low frequency motion \emph{and} conformal distributions has provided new insight into the hierarchy of relationships that may govern the successful and robust function of FMO as an EET device, as quantified by our PCA and correlation results. The important role of low frequency motion in catalysis and other functions, and its selection in evolution has recently become a subject of great interest \citep{marsh2014parallel}, particularly the idea that slow, functional motion is more strongly conserved than other fluctuations and that ensembles of molecular configurations generated in simulation often show strong correlation with the distributions of related structures produced by evolution. These topics, for which the methods presented in this paper are ideally suited to study, will be the subject of forthcoming work.

%
%

\begin{acknowledgements}
We are grateful to Nicholas Hine (University of Cambridge) and Stephen Wells (University of Bath) for helpful discussions.
A.S.F. is supported by a Doctoral Research Award from Microsoft
Research. D.J.C. is supported by a Marie Curie International Outgoing
Fellowship within the Seventh European Community Framework
Programme. A.W.C. is supported by the Winton Programme for the
Sustainability of Physics.
\end{acknowledgements}

\bibliographystyle{spbasic}      
\bibliography{ASF_Paper_bibliography}   

\begin{thebibliography}{70}
\providecommand{\natexlab}[1]{#1}
\providecommand{\url}[1]{{#1}}
\providecommand{\urlprefix}{URL }
\expandafter\ifx\csname urlstyle\endcsname\relax
  \providecommand{\doi}[1]{DOI~\discretionary{}{}{}#1}\else
  \providecommand{\doi}{DOI~\discretionary{}{}{}\begingroup
  \urlstyle{rm}\Url}\fi
\providecommand{\eprint}[2][]{\url{#2}}

\bibitem[{Adolphs and Renger(2006)}]{Adolphs:2006yq}
Adolphs J, Renger T (2006) How proteins trigger excitation energy transfer in
  the {FMO} complex of green sulfur bacteria. Biophys J 91(8):2778--97

\bibitem[{Adolphs et~al(2008)Adolphs, M\"{u}h, Madjet, and
  Renger}]{Adolphs:2008fu}
Adolphs J, M\"{u}h F, Madjet A, Renger T (2008) Calculation of pigment
  transition energies in the {FMO} protein: from simplicity to complexity and
  back. Photosynth Res 95(2-3):197--209

\bibitem[{Anna et~al(2014)Anna, Scholes, and van Grondelle}]{anna2014little}
Anna JM, Scholes GD, van Grondelle R (2014) A little coherence in
  photosynthetic light harvesting. J BioSci 64(1):14--25

\bibitem[{Barzega et~al(2009)Barzega, Moosavi-Movahedi, Pedersen, and
  Miroliaei}]{proline_rigidity}
Barzega A, Moosavi-Movahedi A, Pedersen J, Miroliaei M (2009) Comparative
  thermostability of mesophilic and thermophilic alcohol dehydrogenases:
  Stability-determining roles of proline residues and loop conformations. Enzym
  Microb Technol 45(2):73 -- 79

\bibitem[{Belfield et~al(2014)Belfield, Cole, Martin, Payne, and
  Chau}]{belfield2014}
Belfield WJ, Cole DJ, Martin IL, Payne MC, Chau PL (2014) Constrained geometric
  simulation of the nicotinic acetylcholine receptor. J Mol Graphics Model

\bibitem[{Blankenship(2002)}]{blankenship}
Blankenship RE (2002) Molecular Mechanisms of Photosynthesis. Wiley

\bibitem[{Brixner et~al(2005)Brixner, Stenger, Vaswani, Cho, Blankenship, and
  Fleming}]{exp_EET}
Brixner T, Stenger J, Vaswani HM, Cho M, Blankenship RE, Fleming GR (2005)
  Two-dimensional spectroscopy of electronic couplings in photosynthesis.
  Nature 434(7033):625--628

\bibitem[{Caruso et~al(2009)Caruso, Chin, Datta, Huelga, and
  Plenio}]{caruso2009highly}
Caruso F, Chin AW, Datta A, Huelga SF, Plenio MB (2009) Highly efficient energy
  excitation transfer in light-harvesting complexes: The fundamental role of
  noise-assisted transport. J Chem Phys 131(10):105,106

\bibitem[{Caruso et~al(2010)Caruso, Chin, Datta, Huelga, and
  Plenio}]{caruso2010entanglement}
Caruso F, Chin AW, Datta A, Huelga SF, Plenio MB (2010) Entanglement and
  entangling power of the dynamics in light-harvesting complexes. Phys Rev A
  81(6):062,346

\bibitem[{Case et~al(2009)Case, Darden, Cheatham, Simmerling, Wang, Duke, Luo,
  Crowley, Walker, Zhang, Merz, Wang, Hayik, Roitberg, Seabra, Kolossv\'{a}ry,
  Wong, Paesani, Vanicek, Wu, Brozell, Steinbrecher, Gohlke, Yang, Tan, Mongan,
  Hornak, Cui, Mathews, Seetin, Sagui, Babin, and Kollman}]{AMBER11}
Case DA, Darden TA, Cheatham TE, Simmerling CL, Wang J, Duke R, Luo R, Crowley
  M, Walker RC, Zhang W, Merz KM, Wang B, Hayik S, Roitberg A, Seabra G,
  Kolossv\'{a}ry I, Wong KF, Paesani F, Vanicek J, Wu X, Brozell SR,
  Steinbrecher R, Gohlke H, Yang L, Tan C, Mongan J, Hornak V, Cui G, Mathews
  DH, Seetin MG, Sagui C, Babin V, Kollman PA (2009) AMBER 11. University of
  California, San Francisco

\bibitem[{Ceccarelli et~al(2003)Ceccarelli, Procacci, and
  Marchi}]{force_field_pig}
Ceccarelli M, Procacci P, Marchi M (2003) An ab initio force field for the
  cofactors of bacterial photosynthesis. J Comput Chem 24(2):129--142

\bibitem[{Chin et~al(2013)Chin, Prior, Rosenbach, Caycedo-Soler, Huelga, and
  Plenio}]{chin2013role}
Chin A, Prior J, Rosenbach R, Caycedo-Soler F, Huelga S, Plenio M (2013) The
  role of non-equilibrium vibrational structures in electronic coherence and
  recoherence in pigment-protein complexes. Nature Phys 9(2):113--118

\bibitem[{Chin et~al(2010)Chin, Datta, Caruso, Huelga, and
  Plenio}]{chin2010noise}
Chin AW, Datta A, Caruso F, Huelga SF, Plenio MB (2010) Noise-assisted energy
  transfer in quantum networks and light-harvesting complexes. New J Phys
  12(6):065,002

\bibitem[{Christensson et~al(2012)Christensson, Kauffmann, Pullerits, and
  Mančal}]{christensson2012origin}
Christensson N, Kauffmann HF, Pullerits T, Mančal T (2012) Origin of
  long-lived coherences in light-harvesting complexes. J Phys Chem B
  116(25):7449--7454

\bibitem[{Cole et~al(2013)Cole, Chin, Hine, Haynes, and Payne}]{cole2013}
Cole DJ, Chin AW, Hine NDM, Haynes PD, Payne MC (2013) Toward ab initio optical
  spectroscopy of the {F}enna--{M}atthews--{O}lson complex. J Phys Chem Lett
  4(24):4206--4212

\bibitem[{Collini et~al(2010)Collini, Wong, Wilk, Curmi, Brumer, and
  Scholes}]{collini2010coherently}
Collini E, Wong CY, Wilk KE, Curmi PM, Brumer P, Scholes GD (2010) Coherently
  wired light-harvesting in photosynthetic marine algae at ambient temperature.
  Nature 463(7281):644--647

\bibitem[{David and Jacobs(2011)}]{david2011}
David C, Jacobs D (2011) Characterizing protein motions from structure. J Mol
  Graphics Model 31:41 -- 56

\bibitem[{Dimitrov and Durrant(2013)}]{dimitrov2013materials}
Dimitrov SD, Durrant JR (2013) Materials design considerations for charge
  generation in organic solar cells. Chem Mater 26(1):616--630

\bibitem[{Engel et~al(2007)Engel, Calhoun, Read, Ahn, Mancal, Cheng,
  Blankenship, and Fleming}]{engel2007}
Engel G, Calhoun T, Read E, Ahn T, Mancal T, Cheng Y, Blankenship R, Fleming G
  (2007) Evidence for wavelike energy transfer through quantum coherence in
  photosynthetic systems. Nature 446(7137):782--786

\bibitem[{Fassioli et~al(2014)Fassioli, Dinshaw, Arpin, and
  Scholes}]{fassioli2014photosynthetic}
Fassioli F, Dinshaw R, Arpin PC, Scholes GD (2014) Photosynthetic light
  harvesting: excitons and coherence. J R Soc Interface 11(92):20130,901

\bibitem[{Fidler et~al(2011)Fidler, Harel, Long, and Engel}]{fidler2011two}
Fidler AF, Harel E, Long PD, Engel GS (2011) Two-dimensional spectroscopy can
  distinguish between decoherence and dephasing of zero-quantum coherences. J
  Phys Chem A 116(1):282--289

\bibitem[{Frank(2012)}]{imp_low_Freq_mode}
Frank J (ed)  (2012) Molecular Machines in Biology. Cambridge University Press

\bibitem[{Fulle et~al(2010)Fulle, Christ, Kestner, and Gohlke}]{FRODA_beats_MD}
Fulle S, Christ NA, Kestner E, Gohlke H (2010) {HIV}-1 {TAR} {RNA}
  spontaneously undergoes relevant apo-to-holo conformational transitions in
  molecular dynamics and constrained geometrical simulations. J Chem Info Mod
  50(8):1489--1501

\bibitem[{Gao et~al(2013)Gao, Shi, Ye, Wang, Hirao, and Zhao}]{QMMM_env}
Gao J, Shi W, Ye J, Wang X, Hirao H, Zhao Y (2013) {QM}/{MM} modeling of
  environmental effects on electronic transitions of the {FMO} complex. J Phys
  Chem B 117(13):3488--3495

\bibitem[{G{\'e}linas et~al(2014)G{\'e}linas, Rao, Kumar, Smith, Chin, Clark,
  van~der Poll, Bazan, and Friend}]{gelinas2014ultrafast}
G{\'e}linas S, Rao A, Kumar A, Smith SL, Chin AW, Clark J, van~der Poll TS,
  Bazan GC, Friend RH (2014) Ultrafast long-range charge separation in organic
  semiconductor photovoltaic diodes. Sci 343(6170):512--516

\bibitem[{Grant et~al(2006)Grant, Rodrigues, ElSawy, McCammon, and
  Caves}]{bio3d}
Grant BJ, Rodrigues APC, ElSawy KM, McCammon JA, Caves LSD (2006) Bio3d: an {R}
  package for the comparative analysis of protein structures. Bioinform
  22(21):2695--2696

\bibitem[{Harel and Engel(2012)}]{harel2012quantum}
Harel E, Engel GS (2012) Quantum coherence spectroscopy reveals complex
  dynamics in bacterial light-harvesting complex 2 ({LH2}). Proc Natl Acad Sci
  USA 109(3):706--711

\bibitem[{Hildner et~al(2013)Hildner, Brinks, Nieder, Cogdell, and
  Hulst}]{robust_energy_disorder}
Hildner R, Brinks D, Nieder J, Cogdell R, Hulst N (2013) Quantum coherent
  energy transfer over varying pathways in single light-harvesting complexes.
  Sci 340(6139):1448--1451

\bibitem[{Huelga and Plenio(2013)}]{huelga2013vibrations}
Huelga S, Plenio M (2013) Vibrations, quanta and biology. Contemp Phys
  54(4):181--207

\bibitem[{Ishizaki and Fleming(2012)}]{good_review}
Ishizaki A, Fleming GR (2012) Quantum coherence in photosynthetic light
  harvesting. Annu Rev Condens Matter Phys 3:333--61

\bibitem[{Jing et~al(2012)Jing, Zheng, Li, and Shi}]{QMMM_rb}
Jing Y, Zheng R, Li H, Shi Q (2012) Theoretical study of the
  electronic--vibrational coupling in the {Q}y states of the photosynthetic
  reaction center in purple bacteria. J of Phys Chem B 116(3):1164--1171

\bibitem[{Jolley et~al(2006)Jolley, Wells, Hespenheide, Thorpe, and
  Fromme}]{jolley2006}
Jolley CC, Wells SA, Hespenheide BM, Thorpe MF, Fromme P (2006) Docking of
  photosystem {I} subunit {C} using a constrained geometric simulation. J Am
  Chem Soc 128:8803--8812

\bibitem[{Knox and Spring(2003)}]{knox03}
Knox RS, Spring BQ (2003) Dipole strengths in the chlorophylls. Photochem
  Photobiol 77(5):497--501

\bibitem[{Kozuska et~al(2014)Kozuska, Paulsen, Belfild, Martin, Cole, Holt, and
  Dunn}]{kozuska2014}
Kozuska JL, Paulsen IM, Belfild WJ, Martin IL, Cole DJ, Holt A, Dunn SMJ (2014)
  Impact of intracellular domain flexibility upon properties of activated human
  5-{HT3} receptors. Br J Pharmacol 171:1617--1628

\bibitem[{Kreisbeck and Kramer(2012)}]{kreisbeck2012long}
Kreisbeck C, Kramer T (2012) Long-lived electronic coherence in dissipative
  exciton dynamics of light-harvesting complexes. J Phys Chem Lett
  3(19):2828--2833

\bibitem[{Lambert et~al(2012)Lambert, Chen, Cheng, Li, Chen, and Nori}]{QB}
Lambert N, Chen Y, Cheng Y, Li C, Chen G, Nori F (2012) Quantum biology. Nature
  Phys 9(1):10--18

\bibitem[{Lee et~al(2007)Lee, Cheng, and Fleming}]{lee2007coherence}
Lee H, Cheng YC, Fleming GR (2007) Coherence dynamics in photosynthesis:
  protein protection of excitonic coherence. Sci 316(5830):1462--1465

\bibitem[{Li et~al(2012)Li, Wells, Jimenez-Roldan, Romer, Zhao, Sadler, and
  O'Connor}]{li2012}
Li H, Wells SA, Jimenez-Roldan JE, Romer RA, Zhao Y, Sadler PJ, O'Connor PB
  (2012) Protein flexibility is key to cisplatin crosslinking in calmodulin.
  Protein Sci 21:1269--1279

\bibitem[{Ma(2005)}]{NMA_prob}
Ma J (2005) Usefulness and limitations of normal mode analysis in modeling
  dynamics of biomolecular complexes. Structure 13(3):373 -- 380

\bibitem[{Marsh and Teichmann(2014)}]{marsh2014parallel}
Marsh JA, Teichmann SA (2014) Parallel dynamics and evolution: Protein
  conformational fluctuations and assembly reflect evolutionary changes in
  sequence and structure. BioEssays 36(2):209--218

\bibitem[{Metz et~al(2011)Metz, Pfleger, Kopitz, Pfeiffer-Marek, Barringhaus,
  and Gohlke}]{metz2011}
Metz A, Pfleger C, Kopitz H, Pfeiffer-Marek S, Barringhaus KH, Gohlke H (2011)
  Hot spots and transient pockets:\ predicting the determinants of
  small-molecule binding to a protein-protein interface. JChemInfModel
  52:120--133

\bibitem[{Mohseni et~al(2008)Mohseni, Rebentrost, Lloyd, and
  Aspuru-Guzik}]{mohseni2008environment}
Mohseni M, Rebentrost P, Lloyd S, Aspuru-Guzik A (2008) Environment-assisted
  quantum walks in photosynthetic energy transfer. J Chem Phys 129(17):174,106

\bibitem[{M\"{u}h et~al(2007)M\"{u}h, Madjet, Adolphs, Abdurahman, Rabenstein,
  Ishikita, Knapp, and Renger}]{direct}
M\"{u}h F, Madjet M, Adolphs J, Abdurahman A, Rabenstein B, Ishikita H, Knapp
  E, Renger T (2007) $\alpha$--helices direct excitation energy flow in the
  {F}enna--{M}atthews--{O}lson protein. Proc Natl Acad Sci USA
  104(43):16,862--16,867

\bibitem[{Olaya-Castro and Fassioli(2011)}]{Bchl_7_imp}
Olaya-Castro A, Fassioli F (2011) Characterizing quantum-sharing of electronic
  excitation in molecular aggregates. Procedia Chem 3(1):176 -- 184

\bibitem[{Olbrich et~al(2011{\natexlab{a}})Olbrich, Str{\"u}mpfer, Schulten,
  and Kleinekath{\"o}fer}]{correlation}
Olbrich C, Str{\"u}mpfer J, Schulten K, Kleinekath{\"o}fer U
  (2011{\natexlab{a}}) Quest for spatially correlated fluctuations in the {FMO}
  light-harvesting complex. J Phys Chem B 115(4):758--764

\bibitem[{Olbrich et~al(2011{\natexlab{b}})Olbrich, Str{\"u}mpfer, Schulten,
  and Kleinekath{\"o}fer}]{spectral_dens}
Olbrich C, Str{\"u}mpfer J, Schulten K, Kleinekath{\"o}fer U
  (2011{\natexlab{b}}) Theory and simulation of the environmental effects on
  {FMO} electronic transitions. J Phys Chem Lett 2(14):1771--1776

\bibitem[{O'Reilly and Olaya-Castro(2014)}]{o2014non}
O'Reilly EJ, Olaya-Castro A (2014) Non-classicality of the molecular vibrations
  assisting exciton energy transfer at room temperature. Nature Commun 5

\bibitem[{Orengo et~al(1997)Orengo, Michie, Jones, Jones, Swindells, and
  Thornton}]{cath}
Orengo CA, Michie AD, Jones S, Jones DT, Swindells MB, Thornton JM (1997) Cath
  -- a hierarchic classification of protein domain structures. Structure
  5(8):1093 -- 1109

\bibitem[{O’Reilly et~al(2012)O’Reilly, Kolli, Scholes, and
  Olaya-Castro}]{intermediate_rev}
O’Reilly EJ, Kolli A, Scholes GD, Olaya-Castro A (2012) The fundamental role
  of quantized vibrations in coherent light harvesting by cryptophyte algae. J
  Chem Phys 137(17):174,109

\bibitem[{Panitchayangkoon et~al(2010)Panitchayangkoon, Hayes, Fransted, Caram,
  Harel, Wen, Blankenship, and Engel}]{277K}
Panitchayangkoon G, Hayes D, Fransted K, Caram J, Harel E, Wen J, Blankenship
  R, Engel G (2010) Long-lived quantum coherence in photosynthetic complexes at
  physiological temperature. Proc Natl Acad Sci USA 107(29):12,766--12,770

\bibitem[{Plenio and Huelga(2008)}]{plenio2008dephasing}
Plenio MB, Huelga SF (2008) Dephasing-assisted transport: quantum networks and
  biomolecules. New J Phys 10(11):113,019

\bibitem[{Rebentrost et~al(2009)Rebentrost, Mohseni, and
  Aspuru-Guzik}]{QC_role}
Rebentrost P, Mohseni M, Aspuru-Guzik A (2009) Role of quantum coherence and
  environmental fluctuations in chromophoric energy transport. J Phys Chem B
  113(29):9942--9947

\bibitem[{Renger(2008)}]{pda}
Renger G (2008) Primary Processes of Photosynthesis: Principles and Apparatus.
  pt. 1, RSC Publishing

\bibitem[{Renger and M\"{u}h(2013)}]{renger13}
Renger T, M\"{u}h F (2013) Understanding photosynthetic light-harvesting: A
  bottom up theoretical approach. Phys Chem Chem Phys 15:3348--3371

\bibitem[{Renger et~al(2012)Renger, Klinger, Steinecker, Schmidt~am Busch,
  Numata, and M\"{u}h}]{NMA_protein}
Renger T, Klinger A, Steinecker F, Schmidt~am Busch M, Numata J, M\"{u}h F
  (2012) Normal mode analysis of the spectral density of the
  {F}enna--{M}atthews--{O}lson light-harvesting protein: How the protein
  dissipates the excess energy of excitons. J Phys Chem B
  116(50):14,565--14,580

\bibitem[{Rey et~al(2013)Rey, Chin, Huelga, and Plenio}]{rey2013exploiting}
Rey M, Chin AW, Huelga SF, Plenio MB (2013) Exploiting structured environments
  for efficient energy transfer: The phonon antenna mechanism. J Phys Chem Lett
  4(6):903--907

\bibitem[{Roe and Cheatham(2013)}]{matrix}
Roe DR, Cheatham TE (2013) {PTRAJ} and {CPPTRAJ}: Software for processing and
  analysis of molecular dynamics trajectory data. J Chem Theor Comput
  9(7):3084--3095

\bibitem[{Scholes et~al(2011)Scholes, Fleming, Olaya-Castro, and van
  Grondelle}]{scholes2011lessons}
Scholes GD, Fleming GR, Olaya-Castro A, van Grondelle R (2011) Lessons from
  nature about solar light harvesting. Nature Chem 3(10):763--774

\bibitem[{Shim et~al(2012)Shim, Rebentrost, Valleau, and
  Aspuru-Guzik}]{atom_study}
Shim S, Rebentrost P, Valleau S, Aspuru-Guzik A (2012) Atomistic study of the
  long-lived quantum coherences in the {F}enna-{M}atthews-{O}lson complex.
  Biophys J 102(3):649--660

\bibitem[{Skochdopole and Mazziotti(2011)}]{subsystem}
Skochdopole N, Mazziotti DA (2011) Functional subsystems and quantum redundancy
  in photosynthetic light harvesting. J Phys Chem Lett 2(23):2989--2993

\bibitem[{Sun et~al(2008)Sun, Rose, Ananthanarayanan, Jacobs, and
  Yengo}]{sun2008}
Sun M, Rose MB, Ananthanarayanan SK, Jacobs DJ, Yengo CM (2008)
  Characterization of the pre-force-generation state in the actomyosin
  cross-bridge cycle. Proc Natl Acad Sci USA 105:8631--8636

\bibitem[{Tiwari et~al(2013)Tiwari, Peters, and Jonas}]{tiwari2013electronic}
Tiwari V, Peters WK, Jonas DM (2013) Electronic resonance with anticorrelated
  pigment vibrations drives photosynthetic energy transfer outside the
  adiabatic framework. Proc Natl Acad Sci USA 110(4):1203--1208

\bibitem[{Tronrud et~al(1986)Tronrud, Schmid, and Matthews}]{tronrud}
Tronrud DE, Schmid MF, Matthews BW (1986) Structure and x-ray amino acid
  sequence of a bacteriochlorophyll a protein from prosthecochloris aestuarii
  refined at 1.9~\aa{} resolution. J Mol Bio 188(3):443 -- 454

\bibitem[{Van~Amerongen et~al(2000)Van~Amerongen, Valkunas, and
  Van~Grondelle}]{van2000photosynthetic}
Van~Amerongen H, Valkunas L, Van~Grondelle R (2000) Photosynthetic excitons.
  World Scientific

\bibitem[{Wells et~al(2005)Wells, Menor, Hespenheide, and Thorpe}]{bath32820}
Wells S, Menor S, Hespenheide B, Thorpe M (2005) Constrained geometric
  simulation of diffusive motion in proteins. Phys Bio 2(4)

\bibitem[{Wells et~al(2009)Wells, Jimenez-Rolda, and Romer}]{froda_compare}
Wells S, Jimenez-Rolda JE, Romer R (2009) Comparative analysis of rigidity
  across protein families. Phys Bio 6(4)

\bibitem[{Wells(2013)}]{wells-book}
Wells SA (2013) Geometric simulation of flexible motion in proteins. In:
  Livesay DR (ed) Protein Dynamics. Vol. II., Methods in Molecular Biology.,
  vol 1084, Humana Press, New York, pp 173--192

\bibitem[{Wen et~al(2009)Wen, Zhang, Gross, and Blankenship}]{mem_orient}
Wen J, Zhang H, Gross M, Blankenship R (2009) Membrane orientation of the {FMO}
  antenna protein from chlorobaculum tepidum as determined by mass
  spectrometry-based footprinting. Proc Natl Acad Sci USA 106(15):6134--9

\bibitem[{Wolynes(2009)}]{fluctuation_correlation}
Wolynes PG (2009) Some quantum weirdness in physiology. Proc Natl Acad Sci USA
  106(41):17,247--17,248

\bibitem[{Yuen et~al(1980)Yuen, Shipman, Katz, and Hindman}]{quench}
Yuen MJ, Shipman LL, Katz JJ, Hindman JC (1980) {C}oncentration {Q}uenching of
  {F}luorescence from chlorophyll-a, pheophytin-a, pyropheophytin-a and their
  covalently-linked pairs. Photochem and Photobiol 32(3):281--296

\end{thebibliography}

\end{document}